\documentclass[twocolumn]{aastex631}

\usepackage[mathlines]{lineno}
\usepackage{amsmath}
\usepackage{amssymb}
\usepackage{subfigure}
\usepackage{booktabs}
\usepackage{multirow}
\usepackage{url}
\usepackage{color}
\usepackage[T1]{fontenc}
\usepackage{stfloats}
\usepackage{lipsum}

\newcommand\Rg{R_{\rm g}}

\begin{document}

\title{Modeling fast X-ray variability around an accreting black hole}
\author{Yejing Zhan}
\affiliation{School of Astronomy and Space Science, Nanjing University, Nanjing 210093, People’s Republic of China;}

\author{Bei You}
\affiliation{School of Physics and Technology, Wuhan University, Wuhan 430072, People’s Republic of China;}

\correspondingauthor{Bei You}
\email{youbei@whu.edu.cn}

%\author{Phil Uttley}
%\affiliation{Anton Pannekoek Institute, University of Amsterdam, Science Park 904, 1098 XH Amsterdam, The Netherlands}

\author{Adam Ingram}
\affiliation{School of Mathematics, Statistics and Physics, Newcastle University, Herschel Building, Newcastle upon Tyne, NE1 7RU, UK;}

\author{Wenkang Jiang}
\affiliation{Department of Astronomy, School of Physics and Astronomy, Shanghai Jiao Tong University, Shanghai 200240, China;}
\affiliation{Key Laboratory for Particle Astrophysics and Cosmology (MOE), Shanghai 200240, China}

\author{Fayin Wang}
\affiliation{School of Astronomy and Space Science, Nanjing University, Nanjing 210093, People’s Republic of China;}
\affiliation{Key Laboratory of Modern Astronomy and Astrophysics (Nanjing University), Ministry of Education, Nanjing 210093, People's Republic of China;}

\begin{abstract}
X-ray inter-band time lags are observed during the outbursts of black hole X-ray binaries (BHXRBs). Timing analysis of fast variability in low Fourier frequency bands shows that high-energy photons lag behind low-energy photons, a phenomenon referred to as hard lag. Conversely, in high Fourier frequency bands, low-energy photons lag behind high-energy photons, known as soft lag. This frequency-dependent lag spectrum suggests that the lags arise from different physical processes. 
Notably, a trend has been observed wherein the lags shift towards shorter timescales during the rising hard state, indicating an evolution in the inner accretion flow. In this study, we simulate these inter-band lags by conducting Monte Carlo simulations of the rapid variability within the geometry of a jet base corona. We consider both inward propagating accretion rate fluctuations and reverberation (light crossing) delays in our simulations. We successfully reproduce both low-frequency hard lags and high-frequency soft lags in a self-consistent manner.
We replicate the observed evolution of the frequency-dependent lag spectra by varying the geometrical scale of the corona and the viscous frequency of the disc. Finally, we discuss the potential of a spherical corona and emphasize that polarization observations from the Imaging X-ray Polarimetry Explorer (IXPE) and the enhanced X-ray Timing and Polarimetry mission (eXTP) will be crucial for distinguishing the corona's geometry in future studies. 

%X-ray inter-band time lags are observed in the outbursts of black hole X-ray binaries(BHXRBs). Timing analysis of the fast variability in a low Fourier frequency band shows high-energy photons lag low-energy photons(hard lag). In the high Fourier frequency band, low-energy photons are observed lagging high-energy photons(soft lag). The frequency-dependent lag spectrum indicates that the lags originate from different physical processes. A trend of the lags towards shorter timescales in the rising hard state is observed, which suggests an evolution of the inner accretion flow. In this study, we simulate the inter-band lags by conducting Monte Carlo simulations of the fast variability in the geometry of the vertically extended corona, considering both the propagating fluctuation and the reverberation mapping. Low-frequency hard lags and high-frequency soft lags are self-consistently simulated. We successfully reproduced the observed evolution of the frequency-dependent lag spectra, by changing the geometrical scale and viscous frequency of the inner accretion flow. Finally, we discuss the feasibility of the radially extended corona and point out that polarization observations by Imaging X-ray Polarimetry Explorer (IXPE) and enhanced X-ray Timing and Polarimetry mission (eXTP) are required to distinguish the geometry of the corona in the future. 
        
\end{abstract}

% \linenumbers 

\section{introduction}
A black hole X-ray binary (BHXRB) system consists of an ordinary star and a stellar-mass black hole. In these systems, mass from the ordinary star is accreted onto the black hole through stellar wind or Roche lobe overflow, forming an accretion disc. The viscosity within the accretion disc plays a crucial role in enabling the outward transfer of angular momentum. This process allows matter to spiral inward towards the black hole, where it is heated by gravitational energy \citep{shakuraBlackHolesBinary1973}.

% introduce the outburst with long time scale evolution
\par A BHXRB enters an outburst phase due to instability starting from the outer accretion disc \citep{dubusDiscInstabilityModel2001}. During this outburst, BHXRBs efficiently release gravitational energy, resulting in the production of multi-wavelength radiation, with a predominant focus on X-ray radiation \citep{uttleyMultiWavelengthVariability2014, wangMultiwavelengthStudiesXray2022, mandalMultiwavelengthObservationMAXI2024}. These outbursts can last for months or even years, during which significant changes occur in the spectral states and timing properties over a timescale of days \citep{doneModellingBehaviourAccretion2007, huangINSIGHTHXMTObservationsNew2018, buBroadbandVariabilityStudy2021}. Variations in the accretion mode and the corresponding geometry of the inner region of the accretion flow are believed to be contributing factors to long-term variability, allowing for the analysis of the accretion flow's evolution \citep{buissonMAXIJ1820+070NuSTAR2019, wangDiskCoronaJet2021}.

% introduce the short time scale evolution, including board band noise, QPO. Introduce the origin of QPO (models), leading to a geometry origin, then introduce the 
\par The X-ray emission is also known to exhibit fast variability on time scales down to milliseconds \citep{samimiGX3394NewBlack1979,strohmayerMillisecondXRayVariability1996,vanderklisMillisecondOscillationsXray2000}. 
A power density spectrum of the variability is generally composed of a broad-band noise continuum, with a quasi-periodic oscillation (QPO) superimposed on this \citep{ingramReviewQuasiperiodicOscillations2019,zhangEvolutionPhaseLags2017,demarcoProbingBlackHoleAccretion2022,bollemeijerEvidenceDynamicCorona2024,wangAtypicalLowfrequencyQuasiperiodic2024}. Generally, the broad-band noise is thought to arise from fluctuations in the mass accretion rate within the accretion flow \citep[for alternative theories, see][]{lyubarskiiFlickerNoiseAccretion1997, uttleyFluxdependentAmplitudeBroadband2001, arevaloInvestigatingFluctuatingaccretionModel2006}. While the origin of the QPO remains uncertain, the prevailing theory suggests that it results from the relativistic precession of the inner part of the accretion disc \citep{youXRayQuasiperiodicOscillations2018,ingramLowfrequencyQuasiperiodicOscillations2009,xiaoTimingAnalysisSwift2019, shuiPhaseresolvedSpectroscopyLowfrequency2024}.
%The power spectral analysis of such fast variabilities shows a spectrum, characterized by a broadband noise continuum and quasi-periodic oscillations (QPOs)\citep{rapisardaEvolutionHotFlow2014,youXRayQuasiperiodicOscillations2018,deruiterSystematicStudyPhase2019}. 

\par 
%In frequency-resolved timing analysis, fast variability is reflected in milliseconds of lags. 
Hard lags, characterized by high-energy (hard) photons lagging behind low-energy (soft) photons, are typically observed at low Fourier frequencies (a few hertz and below). These hard lags are thought to arise from the propagation of fluctuations in the mass accretion rate from the cold outer regions to the hot inner regions. As a result, soft photons respond more quickly than hard photons \citep{kotovXrayTimelagsBlack2001, ingramModellingVariabilityBlack2012}. 
%In some BHXRBs, 
In all BHXRBs with sufficiently high-quality observations, the lags can exhibit a reversal of sign at high frequencies, resulting in soft lags, which have the opposite definition to hard lags \citep{wangNICERReverberationMachine2022,demarcoObservationsXrayReverberation2019, chandAstroSatViewNewly2022, yuSpectraltimingStudyInner2023}. Soft lags are usually explained by the time delay caused by the reflection process, i.e., reverberation mapping, where the disc reflects hard photons from the hot X-ray source. The reflected photons take a longer path to reach the observer compared to the directly emitted Comptonized photons \citep{emmanoulopoulosGeneralRelativisticModelling2014, uttleyXrayReverberationAccreting2014, wilkinsOriginLagSpectra2013}. Recently, \cite{uttleyLargeComplexXray2024} proposed that the Comptonized photons would heat up the disc to enhance variable disc thermal emission, which could contribute to the soft lags. The zero-crossing point is defined as the frequency at which the soft lag transitions to hard lag. The zero-crossing point at high frequencies is thought to provide insights into the geometry of the inner accretion flow and its evolution \citep{mizumotoXrayShorttimeLags2018, demarcoInnerFlowGeometry2021}. However, there is still no consensus on the relationship between the zero-crossing points and the physical properties of the inner accretion flow. 

%\par During the long time scale variation, lags analysis is     a good probes for the inner accretion flow of the BHXRBs. Generally, the hard lag, which is defined as high energy photons lagging low energy photons, is considered to originate from the time difference due to the propagation of fluctuation of the mass accretion rate from outer cold region to the inner hot region\citep{kotovXrayTimelagsBlack2001,ingramModellingVariabilityBlack2012}. In addition, the photons reflected by the accretion disc must travel a longer distance to reach the observer than the direct emitted Comptonized photons, which is called the reverberation lags. The reverberation lags are responsible for soft lag, which holds an opposite definition of hard lag\citep{emmanoulopoulosGeneralRelativisticModelling2014,uttleyXrayReverberationAccreting2014}. The lags can be easily measured by calculating the cross-spectrum. 

\par In recent years, the analysis of timing properties in BHXRBs has typically revealed the presence and evolution of hard and soft lags, such as in MAXI J1820+070, MAXI J1535-571 and GX 339–4 \citep{demarcoEvolutionReverberationLag2017,karaCoronaContractsBlackhole2019,wangDiskCoronaJet2021,yuSpectraltimingStudyInner2023}. For example, in the outburst of MAXI J1820+070, hard lags are detected in low-frequency bands below several hertz, whereas soft lags dominate higher-frequency bands above ten hertz, with amplitudes ranging from approximately 0.2 to 3 milliseconds. During the rising hard state, a trend for soft lags towards a shorter timescale is observed, where their amplitudes are decreasing \citep{karaCoronaContractsBlackhole2019}, alongside zero-crossing points shifting to higher frequencies \citep{demarcoInnerFlowGeometry2021}. 

\par It is typically thought that hard lags and soft lags arise from viscous processes and reverberation delays \citep{ingramModellingVariabilityBlack2012,ingramExactAnalyticTreatment2013,emmanoulopoulosGeneralRelativisticModelling2014}. The evolution of these lags is believed to be driven by the inner accretion flow \citep{alstonDynamicBlackHole2020,wangNICERReverberationMachine2022,kalemciBlackHolesTiming2022}. Some theoretical models, such as \textsc{Reltrans} \citep{ingramPublicRelativisticTransfer2019}, provide explanations for the observed lag patterns. The \textsc{Reltrans} model assumes a simple lamppost geometry for the hard X-ray source, which illuminates the thin disc with a cutoff power-law spectrum. The "reflection" spectrum re-emitted from the disk features a prominent $\sim 6.4$ keV iron line, a $\sim 20-30$ keV Compton hump, and a strong soft excess. The light crossing delay between direct and reflected X-rays generates a soft lag between the flux in an energy band dominated by the reflection soft excess and the flux in a continuum-dominated band \citep{garciaXRAYREFLECTEDSPECTRA2013}.
% , while the thin disc acts as a soft source that reflects hard photons. Lags patterns are produced by the variability of the power-law spectrum and the delays between hard and soft sources.
However, observational studies imply that the X-ray source is typically radially extended %\citep{haardtTwoPhaseModelXRay1991,esinAdvectionDominatedAccretionSpectral1997,krawczynskiPolarizedXraysConstrain2022,zhangEvolutionCoronaMAXI2022,wangyanan2024} 
\citep{krawczynskiPolarizedXraysConstrain2022,zhangEvolutionCoronaMAXI2022,youObservationsBlackHole2023,wangRmsFluxSlope2024}
    and/or vertically extended \citep{kylafisJetModelGalactic2008,maDiscoveryOscillations2002021,youInsightHXMTObservationsJetlike2021,mendezCouplingAccretingCorona2022}. 
\cite{wilkinsModellingXrayReverberation2016} provided a model that successfully reproduced the low-frequency hard lags and high-frequency soft lags with a corona both radially and vertically extended by assuming an energy descent in the corona and the bulk effect of the corona. Though this model explains the observed lags patterns with reverberation lags, spectral fitting indicates multiple emission components within the low-energy band \citep{demarcoInnerFlowGeometry2021,wangDiskCoronaJet2021,youInsightHXMTObservationsJetlike2021}, suggesting that inter-band lags may result from these multiple components. Thus, a quantitative simulation for frequency-dependent lag spectra that considers an extended corona and self-consistent radiative transfer is necessary. Furthermore, a comprehensive model is essential to explore the connection between the inner accretion flow and the evolution of the lags.

%%% 最后说明重复出来了soft/hard的结果，重点说明得到了演化
\par In this study, we aim to quantitatively simulate the rapid variability of broadband noise in accreting black holes. Our approach will consider not only the propagating fluctuations of the accretion rate within the disc but also the reverberation between the disc and the spatially extended corona. We will implement this research using a Monte Carlo simulation of radiative transfer.
In previous works \citep{youXRayQuasiperiodicOscillations2018, youXRayQuasiperiodicOscillations2020,zhangConstrainingSizeCorona2019}, simulations of radiative transfer were developed for a stable disc, without accounting for fluctuations in the accretion rate. Those simulations included thermal radiation from the disc, the Comptonization within the corona, and the reflection off the disc. By assuming that the corona undergoes Lense-Thirring precession, the simulations successfully explained observed QPO variabilities, such as the energy-dependent fractional root-mean-square (RMS) and variations in the Fe K$\alpha$ line.
In this work, we will improve the simulation code from \cite{youXRayQuasiperiodicOscillations2018, youXRayQuasiperiodicOscillations2020} by incorporating time-dependent fluctuations in the accretion rate as it propagates through the disc. This enhancement will enable us to capture the rapid variability from both the disc and the spatially extended corona.
Furthermore, we will systematically investigate how the viscosity of the inner disc and the disc-corona geometry affect the rapid variability. This analysis may provide insights into the observed evolution of the frequency-dependent lags during the outburst of a BHXRB.
%In \citeauthor{youXRayQuasiperiodicOscillations2018}(\citeyear{youXRayQuasiperiodicOscillations2018},\citeyear{youXRayQuasiperiodicOscillations2020}) 
%Light curves with fast variability in different energy bands are simulated based on the first principle. 

We summarize the physical model, demonstrating the assumed geometry of the corona and the radiative transfer in Sect. \ref{sec:model}. The details of the Monte Carlo simulation are outlined in Sect. \ref{sec:MC}. The numerical results from the simulations are presented in Sect. \ref{sec:result}. In Sect. \ref{sec:discuss}, we will explore the roles of both the delays in propagating fluctuations and the light traveling in the inter-band lag, and deduce the evolution of the accretion flow that accounts for the observed changes in the lags. Additionally, we discuss the potential of a spherical corona in \ref{sec:the_geometry}, to emphasize the importance of the polarization observations.

\section{Model\label{sec:model}}
\par Since the beginning of an outburst, a BHXRB typically exhibits a progression through several states: the rising hard state, where low-frequency hard lags and high-frequency soft lags are observed; the soft state; and the decaying hard state \citep{homanEvolutionBlackHole2005,doneModellingBehaviourAccretion2007,belloniStatesTransitionsBlack2010}. During the rising hard states, X-ray emissions are predominantly characterized by hard X-ray Comptonization occurring in the corona, along with reflection from the disc \citep[see][and references therein]{youInsightHXMTObservationsJetlike2021}. Conversely, in the soft state, the emissions are primarily soft X-ray thermal emissions from the thin disc \citep{remillardXRayPropertiesBlackHole2006}. In this state, the thin disc is generally believed to extend to the innermost stable circular orbit (ISCO), with a weak corona situated above it \citep{youTESTINGWINDEXPLANATION2016}. In the decaying hard state, the thin disc becomes weak enough to be truncated away from the ISCO, leading to the strong corona dominating the X-ray emissions \citep{liuAccretionBlackHoles2022}. Observations indicate that the truncation radius increases as the mass accretion rate decreases, referred to as "disk receding," when the outburst transitions back to a quiescent state \citep{youObservationsBlackHole2023}. However, the geometry of the corona and its evolution during the rising hard state remains a topic of debate. It may be either radially extended \citep{haardtTwoPhaseModelXRay1991,esinAdvectionDominatedAccretionSpectral1997,poutanenNatureSpectralTransitions1997} or vertically extended \citep{kylafisJetModelGalactic2008,maDiscoveryOscillations2002021,pengNICERNuSTARInsightHXMT2024}. Thus, we explore the extended geometries to produce observed timing properties via self-consistent radiative processes.

\subsection{Geometry\label{sec:geometry}}

\begin{figure}[h]
    \centering
    \includegraphics[width=0.9\linewidth]{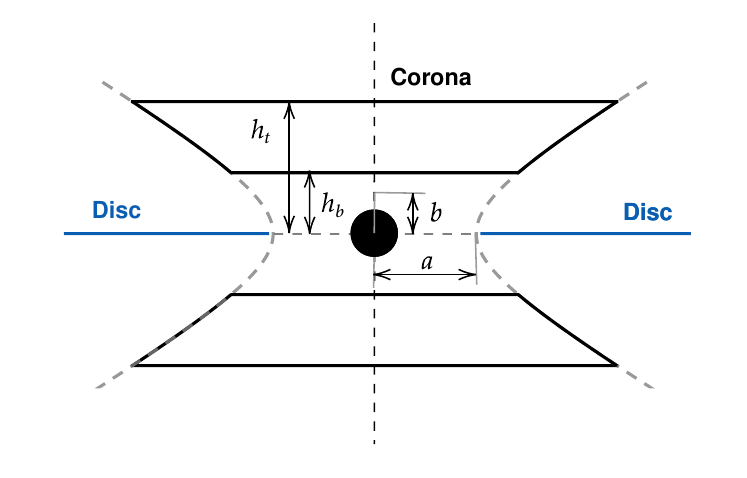}
    \caption{Cross section of the hyperbolic corona floating above the black hole. The upper and lower boundaries are controlled by the height parameters, $h_t,h_b$. The outer surface is a hyperboloid with semi-major axis $a$ and semi-minor axis $b$}
    \label{fig:modelgeo}
\end{figure}

\par We consider a geometrically thin and optically thick disc that extends from the inner radius \( r_{\text{in}} \) to the outer radius \( r_{\text{out}} \). 
%For simplicity, we will first analyze a vertically extended corona, which is enclosed by a hyperbolic surface along with the lower and upper planes (see Fig. \ref{fig:modelgeo}). A radially extended case will be discussed in Sect. \ref{sec:discuss}. 
We mathematically model the corona as a hyperbolic surface with lower and upper surfaces (see Fig. \ref{fig:modelgeo}) , which account for the extended geometry. This geometry is most readily interpreted as the base of the jet, although a more realistic jet base geometry would be parabolic as predicted in the simulation \citep{sridharBulkMotionsBlack2025}. Whether this geometry is vertically or radially extended with respect to the disc depends on the parameters used. The hyperboloid is parameterized by the following equations:
\begin{equation}
\left\{\begin{aligned}
    &\frac{x^2+y^2}{a^2}-\frac{z^2}{b^2}=1\\
    &h_b\leq |z|\leq h_t
\end{aligned}\right.  \label{eq:hyperboloid}
\end{equation}
where $a$ and $b$ are the semi-major and semi-minor axis of the hyperboloid, respectively. $h_t$ and $h_b$ are the top and bottom boundaries in $z$-direction. $\{x,y,z\}$ are the Cartesian coordinates for the black hole frame.  We mainly consider that the corona is wider than it is tall, with the parameters $a=10 \Rg, b=2\Rg$, where $\Rg$ represents gravitational radii. This geometry would ensure a great number of thermal photons from the disc to enter the corona, thereby resulting in the Comptonization emission becoming predominant. An alternative spherical corona case will be discussed in Sect. \ref{sec:discuss}. Some key parameters are shown in Table \ref{tab:explain}, with their definition and default values. 

\subsection{Radiative processes\label{sec:radiative_transfer}}

\par During the rising hard state of an outburst, the X-ray spectrum over a few keV can be described by a power law with a cutoff at $\gtrsim 100\sim 200$ keV \citep{zdziarskiRadiativeProcessesSpectral2004,doneModellingBehaviourAccretion2007}. In the low-energy band, a weak blackbody component is observed in the spectral analysis, together with line emissions\citep{georgeXrayReflectionCold1991,bhattacharyyaEvidenceBroadRelativistic2007}. These findings indicate the presence of multiple emission components arising from radiative processes. Thus, we consider three radiation processes in this work, including thermal radiation by the disc, inverse-Compton scattering from the corona, and the reflection process originating from the hot photons hitting the disc. % with line emissions and a continuum emission.

% only 3 processes are required, thermal, ICS, reflection. modify the statement simulating the fast variability

%In order to simulate light curves with fast variability grounded in the first principle, self-consistent radiative transfer is considered in this job. Generally, our model incorporates three primary radiative mechanisms. First, thermal radiation from the disc contributes to a low-energy spectrum. Second, inverse-Compton scattering(ICS) in high-energy bands, originating from the hot accretion flow, forms a cutoff power-law spectrum. Lastly, reflected Comptonized photons, interacting with the disc, give line emissions below 10 keV and a continuum emission from 10 to 70 keV or a higher energy band. 

%% thermal doesn't need to modify
\par The thermal emission of the thin disc is calculated based on the Novikov-Thorne model \citep{novikovAstrophysicsBlackHoles1973}. Given the black hole mass $M$, the Eddington accretion rate $\dot{m}=\dot{M}/\dot{M}_{\mathrm{Edd}}$, and the dimensionless spin $a$, the model yields the radial emission profile $F(r)$. The temperature profile of the disc $T(r)$ is then calculated following the Stefan-Boltzmann law $T(r)=[F(r)/\sigma]^{1/4}$, where $\sigma$ is the Stefan-Boltzmann constant. The disc emission produces a thermal spectrum in low-energy bands below 1 keV. 

%% move to MC
%Consequently, the disc's thermal radiation establishes a seed soft photon field. The emission position of photons is characterized by the distribution profile $F(r)$, while their energy spectrum adheres to the Planck distribution parameterized by the radial temperature $T(r)$.

%% move to MC
%The frame work of Comptonization is done by \citeauthor{pozdnyakovComptonizationShapingXray1983} (\citeyear{pozdnyakovComptonizationShapingXray1983}) and \citeauthor{goreckiStudyComptonizationRadiation1984}(\citeyear{goreckiStudyComptonizationRadiation1984}), under the assumption of uniform electron density and temperature of the corona. 

\par A fraction of the photons emitted from the disc entering the corona are inverse-Compton scattered by the hot electrons in the corona, transforming these soft photons into hard ones. This process is known as Comptonization \citep{rybickiRadiativeProcessesAstrophysics1991}. Comptonization elevates the thermal spectrum from the disc to a cutoff power-law spectrum in higher energy bands, notably beyond a few keV. The power-law index depends mainly on the optical depth and the temperature of the corona.

\par The disc reflects the Comptonized photons from the corona, yielding characteristic emission lines and continuum emissions. All atomic processes in the disc should be considered, including absorption, excitation of atoms, and emission from the ions in the atmosphere, especially the iron K-shell structure. The shape of the reflected spectra are substantially influenced by the energy spectrum of the incident radiation, as well as the disc's physical conditions—density, ionization, metal abundance, etc. The reflected spectrum consists of emission lines at energies below $\sim 7$ keV, a broad Compton hump peaking at $\sim 20-30$ keV, and a soft excess \citep{georgeXrayReflectionCold1991,dauserIrradiationAccretionDisc2013,garciaXRAYREFLECTEDSPECTRA2013}.

\section{Monte Carlo Simulation\label{sec:MC}}

%\par Many efforts have been made in understanding the origin of the lags \citep{uttleyXrayReverberationAccreting2014,ingramExactAnalyticTreatment2013,uttleyLargeComplexXray2024}. A widely used theoretical model, known as \textsc{Reltrans}, was developed to study the timing properties of BHXRBs \citep{ingramPublicRelativisticTransfer2019}. This model assumes a lamppost geometry for the hard X-ray source. The soft X-ray emission, by contrast, is assumed to originate from a thin accretion disc that reflects these hard photons. In this model, the hard lag is produced by changing the power-law index of the Compton spectrum, and the soft lag is implemented by considering the reverberation of the Compton photons, which depends on the height of the lamppost hard source \citep{mastroserioMultitimescaleXrayReverberation2018}. However, the spectral fitting reveals multiple components in the low-energy band \citep{demarcoInnerFlowGeometry2021,wangDiskCoronaJet2021,youInsightHXMTObservationsJetlike2021}, indicating the inter-band lags may be contributed by these multiple components. Besides, observational studies suggest that the X-ray source tends to be radially extended \citep{haardtTwoPhaseModelXRay1991,esinAdvectionDominatedAccretionSpectral1997,zhangEvolutionCoronaMAXI2022} and/or vertically extended \citep{kylafisJetModelGalactic2008,maDiscoveryOscillations2002021,krawczynskiPolarizedXraysConstrain2022}.  Therefore, in our simulation, we consider three main radiative transfers mentioned in Sect. \ref{sec:radiative_transfer} and an extended corona. 
\par In this simulation, we consider three main radiative processes mentioned in Sect. \ref{sec:radiative_transfer} and an extended corona. We adopt the framework established by \cite{youXRayQuasiperiodicOscillations2018}  concerning radiative transfer. In this study, we introduce fluctuations in the accretion rate that propagate through the disc to account for fast variability. We will simulate three spectral components: the disk component, the Compton component, and the reflection component, in a self-consistent manner. Before introducing temporal variability, we will first outline the Monte Carlo method used for radiative transfer.

% \par Typical lag pattern with a low-frequency hard lag and a high-frequency soft lag infers propagation of fluctuated accretion flow and reverberation mapping in the BHXRBs. Besides, the observed evolution of the amplitude and zero-crossing points in the lag spectra indicates a dynamic transformation of the inner hot flow. To explore the relation between the properties of the inner hot flow and the evolution of the lag, we constructed a Monte Carlo simulation code based on the radiative transfer processes mentioned in Sect. \ref{sec:radiative_transfer}, taking the geometry of the corona, the size of the accretion disc, the parameters of the black hole, etc. as input parameters to produce light curves. The code adopts \citeauthor{youXRayQuasiperiodicOscillations2018}(\citeyear{youXRayQuasiperiodicOscillations2018})'s frame work about radiative transfer, and introduces fluctuation to the accretion rate of the disc to produce fast variability. Three components of the light curves will be self-consistently produced, which are, separately, the disc component, the Compton component, and the reflection component. Before introducing time variability, we first outline the Monte Carlozation for the radiative transfer.

\par The code traces one single photon at a time. The trajectories of the photons are determined geometrically within the Newtonian framework for simplicity, without considering relativistic effects in the present work, e.g., time dilation and the light bending. Time dilation can be described by the gravitational redshift factor (g-factor), defined as the ratio between the frequency of observed and emitted photons. A g-factor of 1 indicates that there is no time dilation. Theoretical calculations in Kerr spacetime demonstrate that g-factor remains around 0.8-1.1 beyond $20R_g$, indicating only mild time dilation relative to flat spacetime \citep{dovciakRadiationAccretionDiscs2004}.
The light bending effect mainly affects the thermal emission and the reverberation process. 
Due to this effect, the radiation from the disc may return to illuminate itself, a phenomenon known as the returning effect. This process modifies the illumination profile to the disc, thereby altering both thermal and reflection spectra. However, the fraction of photons returning to the disc is significant only at very small radii $(<3R_g)$, while this fraction is only $3\%$ at $20 R_g$ for an extremely spinning black hole \citep{dauserEffectReturningRadiation2022}. Spectral fitting further demonstrates minimal differences in parameter estimation when including or excluding the returning effect \citep{huangImpactReturningRadiation2025}. Besides, light bending redistributes coronal illumination, concentrating photons toward the inner region of the disc. Our simulation with a $10R_g$-height lamp-post corona indicates a $\sim 14\%$ increase in illuminated photons within $30 R_g$ and $\sim 8.7\%$ within $50R_g$, compared to the Newtonian case. The caveat is that such an altered illumination profile could influence the traveling distance of the photons and the reflection spectra, particularly when considering the ionization gradient and density gradient. This may modify inter-band time lags \citep{chainakunSimultaneousSpectralReverberation2015}. Moreover, some photons may be gravitationally bound around the black hole, undergoing multiple interactions between the corona and disc. This may form a thermal equilibrium — heating the inner disc while cooling the inner corona — thereby affecting the thermal emission and the Comptonization process. This effect is further complicated and is beyond the scope of this work. Thus, most relativistic effects are negligible; however, some effects remain uncertain regarding their significance. Future models will improve by incorporating these relativistic effects.

% Further studies show that light-bending effects are negligible at distances of tens of from the black hole \citep{dovciakRadiationAccretionDiscs2004}. However, the innermost regions of accretion discs are profoundly influenced by strong gravitational fields. For instance, photons emitted near a black hole with a spin parameter $a = 0.5$ travel with more distance, compared to the case of the spin $a = 0.998$ \citep{wilkinsModellingXrayReverberation2016}. This discrepancy in path length due to light bending induces time lag variations on the order of 0.1 ms. Future models will be improved by incorporating these relativistic effects.}

%The relativistic effect leads mainly to time dilation and light bending. Calculation in Schwarzschild spacetime reveals that gravitational redshift is less than 0.1 at a radius larger than $10\Rg$, which makes a mild time dilation compared to flat spacetime. Research further indicates that the light bending is negligible at tens of $\Rg$ from the black hole \citep{dovciakRadiationAccretionDiscs2004}. But the innermost region of the disc is significantly affected by the gravity. The photons emitted from the innermost region of the disc surrounding a black hole with the spin parameter of 0.5 travel with more distance ($\sim 20 \Rg$) compared to those emitted by a black hole with a spin parameter of 0.998 \citep{wilkinsModellingXrayReverberation2016}. This means the light bending effect leads to a variation of $\sim 10^{-4}\text{ s}$ on time lag.} There would be an improvement in our future model\textbf{, including relativistic effects.

\par Initially, photons with randomly assigned momenta are generated on the accretion disc. Their radially emitting locations are determined using the inverse cumulative distribution function (iCDF), based on the emission profile of the disc \( F(r) \) mentioned in Sect. \ref{sec:radiative_transfer}, combined with a uniform distribution for azimuthal coordinates. The initial energies of the photons follow the blackbody radiation function corresponding to a given disc temperature \( T(r) \). Photons that head directly toward infinity as they leave the disc will be recorded as the observed disk component.

\par A fraction of photons from the disc will strike the corona. These photons may undergo Comptonization, which would contribute to the Compton component. For simplicity, this work does not consider the effects of outflow or inflow velocity, nor the precession of the corona. Notably, the outflow/inflow velocity of the corona would change the seed photon luminosity due to the beaming effects, potentially influencing the simulated results\citep{malzacXraySpectraAccretion2001,wilkinsModellingXrayReverberation2016,youInsightHXMTObservationsJetlike2021}. Additionally, we assume a uniform plasma density and temperature within the corona, for simplicity. A density gradient leads to a gradient on optical depth. So the traveling distance of photons inside the corona would be changed compared to the case with uniform density, resulting in a different spatial lag. Similarly, a temperature gradient results in the energy of escaped photons from the corona being radially or vertically dependent, which would contribute to lags, as demonstrated in \cite{wilkinsModellingXrayReverberation2016}, who assumed a photon index gradient in the corona. The optical depth of the corona is set in the range of 1-3, which ensures the spectrum is hard.

The simulation of Comptonization follows the standard methods established by \cite{pozdnyakovComptonizationShapingXray1983} and \cite{goreckiStudyComptonizationRadiation1984}, which are parameterized by several factors related to the properties of the corona, such as optical depth and electron density, to determine whether the photons will be scattered.

\par The accretion disc reflects photons emitted by the corona, contributing to the reflection component of the spectrum. The reflection spectra of a neutral accretion disc have been calculated self-consistently using Monte Carlo simulations, as detailed in \cite{youXRayQuasiperiodicOscillations2018,youXRayQuasiperiodicOscillations2020}. However, spectral analysis during the rising hard state indicates that the accretion disc must be highly ionized \citep{mastroserioXrayReverberationMass2019,youInsightHXMTObservationsJetlike2021,lucchiniVariabilityPredictorHardtosoft2023}, which significantly alters the reflection spectrum compared to that of a neutral disc. Therefore, a more refined simulation of reflection spectra is necessary, although this will be quite time-consuming.

Instead, we employ the reflection model developed by \cite{garciaXRAYREFLECTEDSPECTRA2013} and \cite{dauserIrradiationAccretionDisc2013}, who created numerical methods to solve the radiation transfer equation, the ionization equilibrium, and energy conservation, incorporating new atomic data. This model assumes plane-parallel geometry and azimuthal symmetry. Given the parameters of the disc, the model produces the normalized reflection spectra, including both continuum and line emission. Details of the calculation will be discussed in Sect. \ref{sec:reflection}.

\par Variabilities have not yet been introduced in the photon generation process. To produce variability, we require information about the timing of the photons. To do so, the simulation begins with a loop of time bins defined by a specified time resolution and duration. Within each time bin, a predetermined number \( N \) of photons is generated. We assume that fast variability stems from fluctuations in the accretion rate, which subsequently affects the weight of each photon. Instead of counting the number of photons, we measure the weight of each photon as a representation of photon flux to introduce fast variability. The details of this approach will be discussed in the next section.

\subsection{Implementation of disc variability\label{sec:fluctuations}}

In this study, we propose that the rapid variability originates from fluctuations in the accretion rate of the disc. 
%This is based on the premise that changes in the accretion rate directly influence the emissions from the disk \citep{shakuraBlackHolesBinary1973}. 
To model the fluctuating accretion flow, we utilize the fluctuation model proposed by \cite{ingramPhysicalModelContinuum2011}, which addresses local fluctuations and their propagation.
The fluctuations propagate inward from the outer radius \( r_{\text{out}} \) to the inner radius \( r_{\text{in}} \) through the accretion flow, merging with local fluctuations along the way. For simplicity, we divide the accretion flow into \( N_{\rm r} \) radial bins, each centered at \( r_n \) in a logarithmic scale. The power spectrum of the local fluctuation at radius \( r_n \) is represented by a zero-centered Lorentzian function:
\begin{equation}
    |{\dot M}(r_n,f)|^2 \propto \frac{1}{1+(f/f_{\mathrm{visc}}(r_n))^2}
    \label{eq:periodogram}
\end{equation} 
\noindent where $f$ is frequency, $f_{\mathrm{visc}}(r_n)=f_{\max}(r_n/r_i)^{-3/2}$ is the viscous frequency at radius $r_n$, in which $r_i$ is the inner edge of the disc \citep{shakuraBlackHolesBinary1973}. $f_{\max}$ is a free model parameter that sets the maximum viscous frequency in the disc. 

\par Based on the randomized method proposed by \cite{timmerGeneratingPowerLaw1995}, the local fluctuation \(\dot{m}(r_n, t)\) in the time domain at radius \(r_n\) is generated by the inverse Fourier transformation of the power spectrum, as stated in equation (\ref{eq:periodogram}), using Gaussian random phases. When considering a flow that propagates from the outer radius \(r_l\) to the inner radius \(r_n\), the fluctuation in each ring should be the superposition of the local fluctuation \(\dot{m}(r_n, t)\) and the fluctuations propagated from the outer rings \citep{ingramExactAnalyticTreatment2013}. This relationship can be expressed as:  
\begin{equation}
    \dot m_n(t)=\prod_{l=1}^n \dot m(r_n,t-\Delta t_{ln})
\end{equation}
\noindent where we use a subscript $n$ to represent the fluctuation in the $n$-th radial bin to distinguish from the local fluctuation and $\Delta t_{ln}$ is the propagation delay between radius $r_l$ and $r_n$, which is naturally given by 
\begin{equation}
    \Delta t_{ln}=\sum_{q=l+1}^n\frac{\mathrm d r_q}{r_q f_{\mathrm{visc}(r_q)}}
\end{equation}
\noindent where $\mathrm d r_n=r_n-r_{n-1}$. Thus, for example, $\Delta t_{nn}=0$ represents the local fluctuation, and $\Delta t_{(n-1)n}$ is the delay from the previous radius. 

\par The varying accretion rate affects the temperature of the accretion disc, which in turn influences the number of emitted photons. The effective temperature of the accretion disc, denoted as \( T_{\text{eff}} \), is proportional to the quarter power of the accretion rate \( [\dot{m}_n(t)]^{1/4} \) \citep{shakuraBlackHolesBinary1973}. Additionally, the number of photons, \( N_p \), emitted from a specific area is proportional to the third power of the effective temperature, \( T^3 \). This leads to a relation where \( N_p \propto [\dot{m}_n(t)]^{3/4} \).
By standardizing the mean values of the fluctuation at each radius to 1, we can approximate \( \dot{m}_n(t) \) as the fluctuated "photon number", which we will refer to as the \textit{weight}. For instance, when there is no fluctuation, the disc emits one photon at a specific time \( t_0 \) and radius \( r_0 \). However, with a fluctuation such that \( \dot{m}_{r_0}(t_0) = 1.1 \), the disc instead emits 1.1 photons. This conceptually means that the number of emitted photons is weighted by \( \dot{m}_n(t) \).
Thus, the fluctuated photon flux within an energy band can be calculated by summing all the weights, which can be expressed mathematically as:
\begin{equation}
     f(E,t)=\sum ^{\mathcal N(E,t)}_q \dot m_{n_q} (t-\tau_q),
     \label{eq:flux}
\end{equation}
where $\mathcal N(E,t)$ is the number of observed photons at time $t$ in energy band $E$, $n_q$ represents the index of radius bin of the $q$-th photon, and $\tau_q$ is the traveling delay from the source to the observer. Our simulation strategy maintains a steady number of observed photons. Consequently, the rapid variability of the system arises from fluctuations in the weights assigned to different spectral components. These weights indicate variations in travel delay, denoted as \(\tau\), and propagation delay, represented by \(\Delta t\), across the energy bands. This variability contributes to the observed patterns in the lag spectra. The dependence on these two types of delays in Equation \ref{eq:flux} suggests that the lags consist of both a spatial component and a viscously propagating component. This preliminary analysis paves the way for a more detailed discussion regarding the physical significance of the lags in Sect. \ref{sec:discuss}.

\subsection{Implementation of disc reflection\label{sec:reflection}}

\par Partial Comptonized photons from the corona are reflected by the disc, resulting in both emission line and continuum emission. In the works of \cite{youXRayQuasiperiodicOscillations2018,youXRayQuasiperiodicOscillations2020}, the reflection for a neutral disc was simulated to study the QPO variation. However, previous studies on BHXRBs have suggested the presence of an ionized disc \citep{youInsightHXMTObservationsJetlike2021}. It has been shown that the flux in the soft band, for instance, 0.1-1 keV, is highly dependent on the ionization state of the disc \citep{garciaXRAYREFLECTEDSPECTRA2013}. Therefore, simulating the reflection of an ionized disc is essential. This involves modeling various processes, such as absorption, excitation, and photoionization, which can be quite time-consuming. To address this, we adopt the table model \textsc{Xillver} \citep{garciaXRAYREFLECTEDSPECTRA2013} to simulate disc reflection.

\begin{figure}[h]
    \centering
    \includegraphics[width=0.9\linewidth]{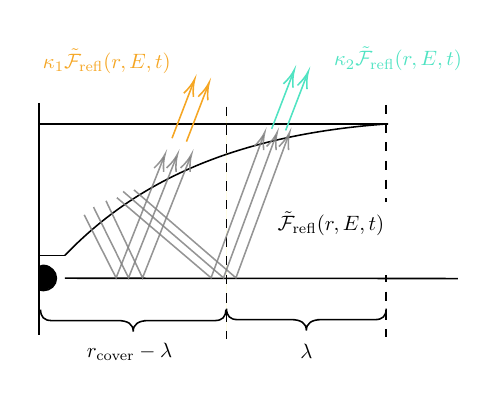}
    \caption{Schematic of the two-zone method. The corona is separated into two zones according to an uncovering length $\lambda$, with transmittance $\kappa_1 $ and $\kappa_2 $, respectively. Denoting the radius of the vertical projection of the corona on the disc by $r_{\text{cover}}$ and the reflection emission $\mathcal{F}(r,E,t)$ at radius $r$, when $r<r_\text{cover}-\lambda$, $\mathcal{F}_\text{refl}(r,E,t)=\kappa_1 \tilde{\mathcal{F}}_\text{refl}(r,E,t)$ and when$r_\text{cover}-\lambda<r<r_\text{cover}$, $\mathcal{F}_\text{refl}(r,E,t)=\kappa_2 \tilde{\mathcal{F}}_\text{refl}(r,E,t)$.}
    \label{fig:shielding-effect}
\end{figure}

\par The \textsc{Xillver} model relies on several discrete parameters, the most important of which are the photon index $\Gamma$, the ionization parameter $\xi$, the iron abundance $A_{\text{Fe}}$, and the angle of inclination $i$ Together, these parameters create a discrete parameter space, and the shape of the reflection spectra can be generated by interpolating within this space. Additionally, the amplitude of the reflection spectra is normalized by the integral of the incident flux $\mathcal{F}_X(E)$ over the range from 0.1 keV to 1000 keV.
\begin{equation}
    \int^{1000\mathrm{keV}}_{0.1\mathrm{keV}}\mathcal F_X(E)\mathrm{d}E=10^{-20}\frac{n_0\xi_0}{4\pi}
    \label{eq:norm}
\end{equation}
where $n_0=10^{15}\mathrm{~cm^{-3}}$ and $\xi_0=1\mathrm{~erg~cm~s^{-1}}$ \citep{dauserNormalizingRelativisticModel2016}. Therefore, the shape of the reflection spectra depends on the parameters, and the amplitude depends on the integral of the incident flux.

\par The outgoing hard photons from the corona to the disc are collected in different time and radius bins as the incident photon flux, denoted as \(\mathcal{N}(r,E,t)\). This flux follows the relationship \(\mathcal{N} \propto E^{-\Gamma}\), which connects the photon flux \(\mathcal{N}\) to the photon index \(\Gamma\). By fitting the energy spectra output by the Monte Carlo code, we produce a function \(\Gamma(r,t)\). This function is then averaged over both \(r\) and \(t\) under the assumptions of an isotropic incident flux and steady Compton radiation, establishing the averaged photon index \(\Gamma\) as a key parameter for the reflection spectra.
Given the disc configuration parameters \(\xi\), \(A_{\text{Fe}}\), along with the fixed observation angle denoted by \(i\) and the averaged \(\Gamma\), we determine the shape of the reflection spectra \(\mathcal{F}_{\text{shape}}(E)\) by interpolating from a tabulated model.
%The interpolation is then done in a spectral-fitting software \textsc{Xspec} \citep{arnaud1996xspec}.

\par To obtain the actual amplitude of the reflection spectrum in relation to the incident spectrum, a coefficient is needed to adjust for the normalization of the table model. This can be expressed as $\tilde{\mathcal{F}}_{\text{refl}}(r,E,t)=c(r,t)\mathcal{F}_{\text{shape}}(E)$, where $\tilde{\mathcal{F}}_{\text{refl}}$ represents the actual reflected flux. Assuming a linear relation between the integral of incident flux and that of reflected flux, 
% \begin{widetext}
\begin{equation}
\begin{aligned}
    \frac{\text{actual incident flux}}{\text{normalized incident flux}}=\frac{\int^{1000\mathrm{keV}}_{0.1\mathrm{keV}}\mathcal F(r,E,t)\mathrm d E}{\int^{1000\mathrm{keV}}_{0.1\mathrm{keV}}\mathcal F_X(E)\mathrm d E}=\\
    \frac{\text{actual reflected flux}}{\text{normalized reflected flux}}=\frac{\int \tilde{\mathcal{F}}_{\text{refl}}(r,E,t) \mathrm dE}{\int \mathcal F_{\text{shape}}(E)\mathrm d E}
\end{aligned}
    \label{eq:linear-norm}
\end{equation}
% \end{widetext}
\noindent where the definition of the integral of the normalized incident flux $\mathcal F_X(E)$ is shown in Equation (\ref{eq:norm}). Therefore, the coefficient to compensate for the normalization is 
\begin{equation}
    c(r,t)=4\pi\times 10^5 \int^{1000\mathrm{keV}}_{0.1\mathrm{keV}}\mathcal N(r,E,t)E~\mathrm d E
    \label{eq:anti-norm}
\end{equation}
where the incident flux $\mathcal F(r,E,t)$ is expressed in terms of the photon flux $\mathcal{N}(r,E,t)E$.

\par Given the geometric constraint where a portion of the disc is obscured by the corona, it is important to consider the shielding effect. To address this comprehensively, we can use a self-consistent method to calculate the shielding projection from the observer's perspective. This approach allows us to determine the transmittance of photons traveling through the corona from each point on the disc. However, due to the computational complexity and time required for this method, we will introduce an approximation.
In this approximation, we divide the corona into two zones with different transmittance values, denoted as \(\kappa_1\) and \(\kappa_2\). We also define a radius, \(r_{\text{cover}}\), which is defined as the vertical projection radius of the corona on the disc, $r_\text{cover}=a\sqrt{1+(h_t/b)^2}$, and a parameter free \(\lambda\), defined as the \textit{uncovering length}, to quantify the shielding effect (see Fig. \ref{fig:shielding-effect}).
Consequently, for regions where the radius \(r\) is less than \(r_{\text{cover}} - \lambda\), the reflected flux is influenced by the transmittance \(\kappa_1\), resulting in the expression \({\mathcal{F}_\text{refl}}(r,E,t) = \kappa_1 \tilde{\mathcal{F}}_\text{refl}(r,E,t)\). In cases where \(r_{\text{cover}} - \lambda < r < r_{\text{cover}}\), the flux is adjusted to \(\mathcal{F}_\text{refl}(r,E,t) = \kappa_2 \tilde{\mathcal{F}}_\text{refl}(r,E,t)\). $\kappa_1$, $\kappa_2$ and $\lambda$ will be specified as model parameters.

\par The two-zone method simplifies a complex three-dimensional reflection problem into a straightforward two-dimensional one. This approach avoids the need to calculate the actual covered area from the observer's perspective and eliminates the simulation of transmittance for covered radii. While we do neglect variations in axial and transmittance factors, which may introduce some degree of error, we opted for the two-zone method due to computational limitations. Future work will focus on improving this method.

\par Emission from all radii contributes to the observed flux. Therefore, confirming both the amplitude and the shape, the actual observed reflection spectra, including the shielding effect, is 
\begin{equation}
\begin{aligned}
    \mathcal F_{\text{refl}}(t,E)&=\sum_r \mathcal  F_\text{refl}(r,E,t)\\
    &=\sum_{r<r_\text{cover}-\lambda}\kappa_1 c(r,t)\mathcal F_{\text{shape}}(E)\\
    &+\sum_{r_\text{cover}-\lambda<r<r_\text{cover}} \kappa_2 c(r,t) \mathcal F_{\text{shape}}(E)\\
    &+\sum_{r>r_\text{cover}} c(r,t) \mathcal F_{\text{shape}}(E)
\end{aligned}
\end{equation}

\section{Result \label{sec:result}}

We aim to simulate the inter-band lags across Fourier frequencies in the BHXRB. This will be achieved through Monte Carlo simulations of the rapid variability within the disc-corona system, taking into account both propagating fluctuations and reverberation mapping. Our goal is to quantitatively test whether these propagating fluctuations and reverberation mapping can explain the observed hard lag at low frequencies and the soft lag at high frequencies. Additionally, we will compare the results of our simulations with the observed evolution of frequency-dependent lags, which will help us constrain the geometry of the accretion flow and its evolution during outbursts. In this work, we will conduct simulations and comparisons specifically for MAXI J1820+070, which has provided valuable observations of fast variability \citep{karaCoronaContractsBlackhole2019, demarcoInnerFlowGeometry2021,kawamuraMAXIJ1820+070Xray2023}.

\par To simulate rapid variability, we introduce disc variability by incorporating the propagation of fluctuations into the simulation described in \citep{youXRayQuasiperiodicOscillations2018}, which modeled radiative transfer in the disc and corona. We assume a black hole mass of \( M = 7 \, M_\odot \), an Eddington-scaled mass accretion rate of \( \dot{m} \equiv \dot{M} / \dot{M}_{\text{Edd}} = 0.01 \), and a disc temperature in the innermost region of \( T_i = 0.15 \, \text{keV} \). We set the inner radius of the disc to \( r_{\text{in}} = 10 \, R_g \), and the outer radius to \( r_{\text{out}} = 500 \, R_g \).  Additionally, we assume a viscous factor of \( \alpha = 0.1 \) and a hot flow temperature of \( 100 \, \text{keV} \).
The parameters for disc reflection are as follows: we set the ionization parameter to \( \xi = 3.8 \), and the logarithmic radial bins are divided into \( N_r = 20 \) segments. The iron abundance is set to \( 2 \), and the angle of inclination is \( i = 60^\circ \). We define the uncovering length in the two-zone method as \( \lambda = h_t \tan(i) \), where \( h_t \) represents the upper height of the corona. For the transmittance, we set \( \kappa_1 = 0 \) and \( \kappa_2 = 1 \) empirically.

\begin{table*}[htbp]
\centering
\caption{Some key parameters in this model and their definitions}\label{tab:explain}
\begin{tabular}{@{}lclllcl@{}}
\toprule
 & notation &  & definition & &default value& unit \\ \midrule
 & $a$ &  & the semi-major axis of the hyperbolic corona& & 10& $R_g$ \\
 & $b$ &  & the semi-minor axis of the hyperbolic corona && 2& $R_g$ \\
 & $h_t$ &  & the height of the top boundary of the corona & & 30&$R_g$ \\
 & $h_b$ &  & the height of the bottom boundary of the corona && 6& $R_g$ \\
 & $f_\text{max}$ &  & the maximum viscous frequency of the disc &&4 & Hz \\
 & $i$ &  & the inclination angle with respect to the normal to the disc &&60 & $^\circ$ \\
 & $\kappa_1,\kappa_2$ &  & the transmittance used in shielding effect of reflection& &0,1 &  \\
 & $\lambda$ &  & the uncovering length used in shielding effect of reflection && $h_t\tan (i)$& $R_g$ \\
 & $r_\text{cover}$ &  & the projection of corona &&$a\sqrt{1+(h_t/b)^2}$ &$R_g$ \\ \bottomrule
\end{tabular}
\end{table*} 

\par The hyperbolic geometry of the corona is characterized by a semi-major axis \( a = 10 \, \Rg \) and a semi-minor axis \( b = 2 \, \Rg \). The lower boundary is set at \( h_b = 6 \, \Rg \), while the upper boundary is at \( h_t = 30 \, \Rg \). The viscosity of the inner hot flow is determined by the maximum viscous frequency \( f_{\text{max}} = 4 \, \text{Hz} \). We simulated light curves for three components: the disc, the Compton component, and the reflection component. These simulations were conducted over a duration of 200 seconds, with a time resolution of 1/1000 second. 
\begin{figure*}[h]
    \centering
    \includegraphics[width=1\linewidth]{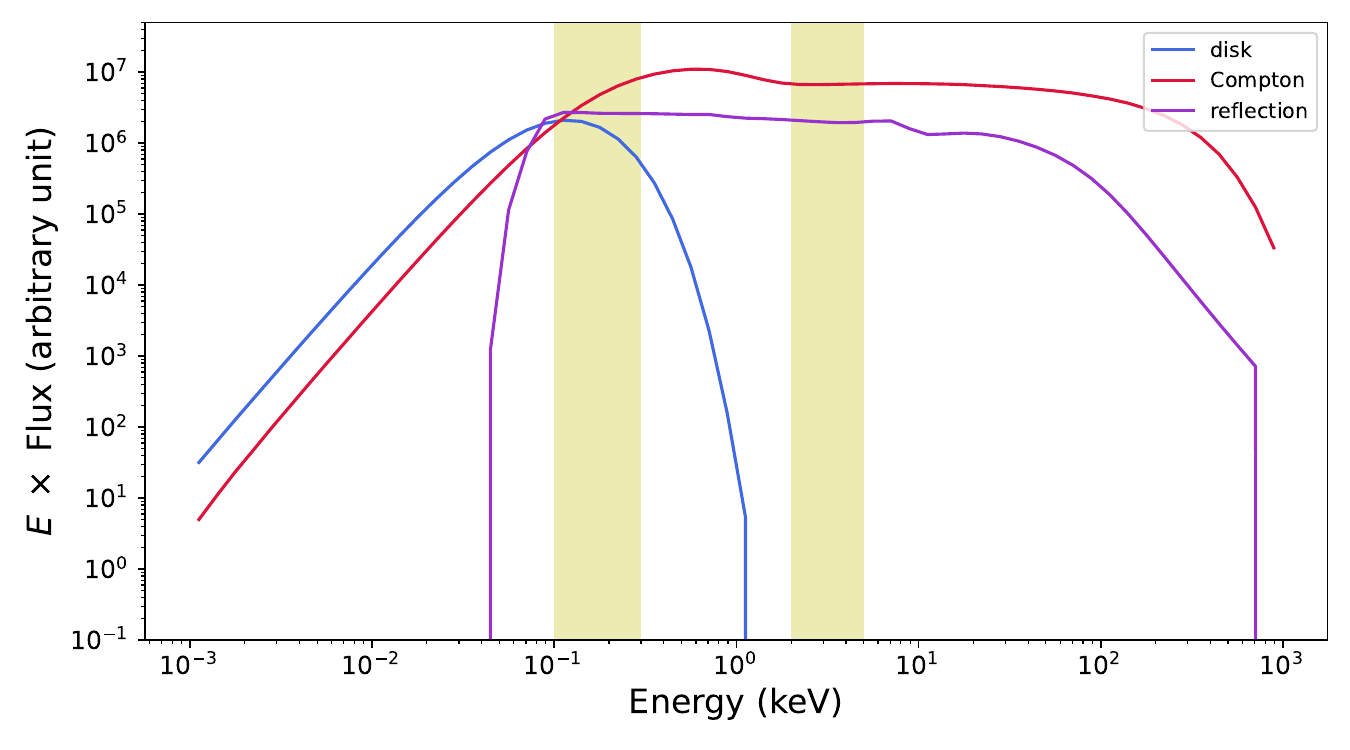}
    \caption{The time-averaged energy spectrum of the three components. The Y-axis is labeled by the energy multiplied by the flux by convention. The energy bands 0.1-0.3 keV and 2-5 keV are selected as two bands in the timing analysis, colored in the yellow block.}
    \label{fig:Espec}
\end{figure*}

\par
\begin{figure*}[htbp]
   \centering
    \includegraphics[width=1\linewidth]{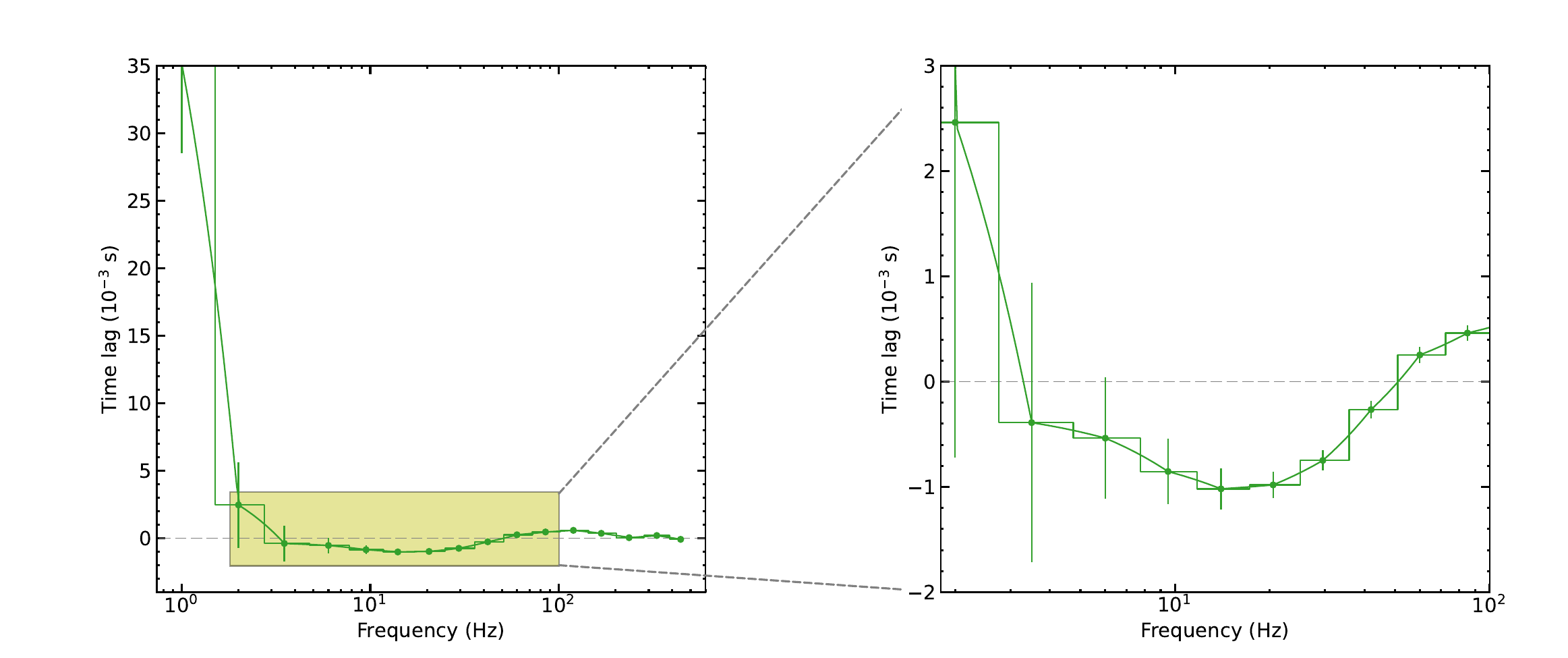}
   \label{fig:lag-spectra-single}
   \caption{The left figure presents the frequency-lag spectrum, calculated between the energy ranges of 0.1-0.3 keV and 2-5 keV. The right figure illustrates a more detailed view of this spectrum, highlighting a specific window that corresponds to the colored section in the left figure. Both low-frequency hard lag and high-frequency soft lag are presented in the results. The solid line in the figures is generated through linear interpolation. The analysis was conducted using the \textsc{Python} package \textsc{Stingray}.}
\end{figure*}

 The time-averaged energy spectrum of the three components is presented in Fig. \ref{fig:Espec} for demonstration purposes. The spectrum exhibits a cutoff power-law component with photon index $\sim 2$, a thermal component with the peak temperature $\sim 0.1$ keV, and a high-ionized reflection component. These features are consistent with the spectral fitting results, e.g., \cite{youInsightHXMTObservationsJetlike2021} and \cite{gaoLowfrequencyQuasiperiodicOscillation2023}. We focus on an energy band of interest ranging from 0.1 to 0.3 keV, where all components are comparable, and a reference band from 2 to 5 keV, where the Compton component is prominent. The lag spectra between these two bands can be calculated using the Python package \textsc{Stingray} \citep{huppenkothenStingrayModernPython2019}, as shown in Fig. \ref{fig:lag-spectra-single}. The errors arise from the incoherent components between the two bands and the stochastic nature of the variability \citep{bendatRandomDataAnalysis2011,uttleyXrayReverberationAccreting2014}. The solid line in the plot represents a linear interpolation to guide the eye. We observe a low-frequency hard lag and a high-frequency soft lag in the spectrum, which are consistent with observations\citep {karaCoronaContractsBlackhole2019,yuSpectraltimingStudyInner2023,demarcoInnerFlowGeometry2021}.

% \begin{figure}[h]
%    \centering
%    \subfigure[\label{fig:lag-full-single}]{\includegraphics[width=0.45\linewidth]{single_xillver.pdf}}\hspace{20pt}
%    \subfigure[\label{fig:lag-windowed-single}]{\includegraphics[width=0.45\linewidth]{single_windowed_xillver.pdf}}
%    \label{fig:lag-spectra-single}
%    \caption{The analysis was conducted using the \textsc{Python} package \textsc{Stingray}. Figure (a) presents the time-averaged frequency-lag spectrum, calculated between the energy ranges of 0.1-0.3 keV and 2-5 keV, with an averaged segment size of 1 second. Figure (b) illustrates a more detailed view of this spectrum, highlighting a specific window that corresponds to the colored section in Figure (a). Both low-frequency hard lag and high-frequency soft lag are presented in the results. The solid line in the figures is generated through linear interpolation.}
% \end{figure}

\par
\begin{figure*}[htbp]
   \centering
   \includegraphics[width=1.\linewidth]{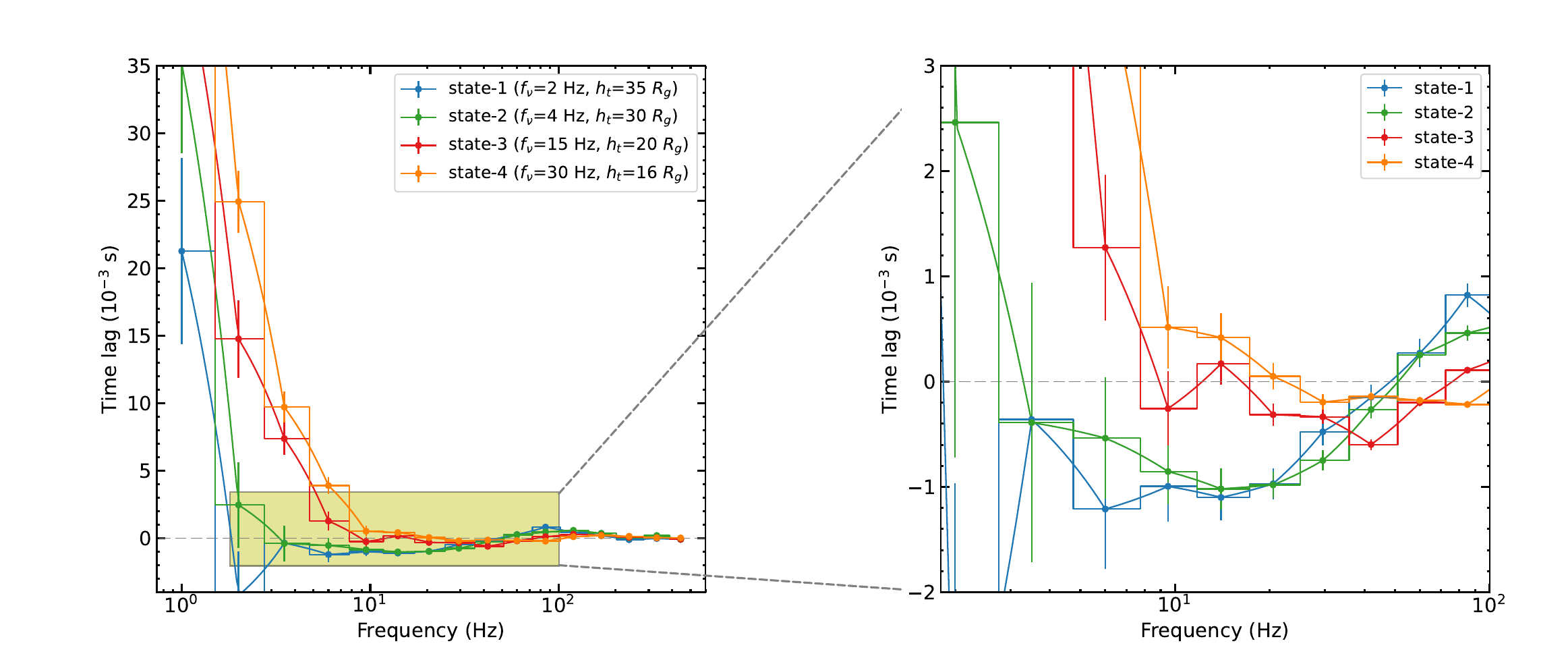}
   \label{fig:lag-spectra}
   \caption{The left figure is the lag spectra between 0.1-0.3 keV and 2-5 keV as the model parameters change. The right figure provides a detailed view, featuring a window highlighted in the left figure. The figures illustrate a scenario where the upper height decreases from $35~\Rg$ to $16~\Rg$, while the lower height remains fixed at $6~\Rg$. During this process, the maximum viscous frequency increases from 2 Hz to 30 Hz. The amplitude of the soft lags diminishes, and the lag pattern shifts to a higher frequency from state 1 to state 4. The solid line indicates linear interpolation to guide the eye. The parameters in each state are listed in the legend.}
\end{figure*}

 A trend towards shorter timescales for the lags is observed in MAXI J1820+070 as the accretion rate increases \citep{karaCoronaContractsBlackhole2019}. Moreover, \cite{demarcoInnerFlowGeometry2021} note that the movement of the zero-crossing points from low to high frequencies indicates an evolution of the physical properties of the inner accretion flow \citep{alstonDynamicBlackHole2020, wangNICERReverberationMachine2022, kalemciBlackHolesTiming2022}. We attribute these physical properties to the geometry of the corona and the inner viscosity. Our simulations captured the evolution of the corona's geometrical size and viscosity frequency, with the corona's upper bound decreasing from $35~\Rg$ to $16~\Rg$, accompanied by an increase in the maximum viscous frequency from 2 Hz to 30 Hz. Four evolving states are simulated, labeled state-1 to state-4, where state-1 is characterized by a maximum viscous frequency of 2 Hz and an upper bound for the corona of $35~\Rg$. The lag spectra between 0.1-0.3 keV and 2-5 keV for the four states are presented in Fig. \ref{fig:lag-spectra}.

% \begin{figure*}[h]
   
%    \subfigure[\label{fig:lag-full}]{\includegraphics[width=0.45\linewidth]{dcnr_whole_lag_xillver.pdf}}\hspace{20pt}
%    \subfigure[\label{fig:lag-windowed}]{\includegraphics[width=0.45\linewidth]{dcnr_whole_lag_windowed_xillver.pdf}}
%    \label{fig:lag-spectra}
%    \caption{Figure (a) shows the lag spectra between 0.1-0.3 keV and 2-5 keV as the model parameters changes. Figure (b) provides a detailed view, featuring a window highlighted in Figure (a). The figures illustrate a scenario where the upper height decreases from $35~\Rg$ to $16~\Rg$, while the lower height remains fixed at $6~\Rg$. During this process, the maximum viscous frequency increases from 2 Hz to 30 Hz. The amplitude of the soft lags diminishes, and the lag pattern shifts to a higher frequency from state 1 to state 4. The solid line indicates linear interpolation to guide the eye. The parameters in each state are listed in Tab. \ref{tab:para}.}
% \end{figure*}

%\par The configuration of the corona is fixed by the hyperbolic boundary with the semi-major axis $a_p=10\Rg$, the semi-minor axis $b_p=2\Rg$. The lower boundary $h_b$ is a constant of $6\Rg$ over time. Since a specific evolution of lags is observed, which is believed to originate from the inner region of the accretion flow \citep{alstonDynamicBlackHole2020,wangNICERReverberationMachine2022,kalemciBlackHolesTiming2022}, our simulations captured the evolution of the corona's geometry and viscosity frequency - with the corona's upper bound reduced from $35\Rg$ to $16\Rg$, together with a growth in the maximum viscous frequency from 2Hz to 30Hz. Four evolving states are simulated, marked by time 1 to time 4, where time 1 is simulated with a maximum viscous frequency of 2Hz and an upper bound of the corona of $35\Rg$. 

\par As the height of the corona decreases and the maximum viscous frequency increases, the amplitude of the soft lag diminishes, while the zero-crossing points shift to higher frequencies. This pattern of evolution aligns closely with the observations, suggesting a connection between the changes in the inner accretion flow and the evolution of the lags. This will be further discussed in Section \ref{sec:discuss}.

% We investigate the effects of geometrical changes by maintaining the maximum viscous frequency at 4 Hz while reducing the height of the corona from \(35 ~\Rg\) to \(16 ~\Rg\). As shown in Fig. \ref{fig:vgeo}, when the frequency exceeds 4 Hz, the amplitude of the lag decreases, and the zero-crossing point for the soft lag shifts monotonically to higher frequencies. 

% On the other hand, we also examine the effects of viscosity by fixing the corona height at $30 \Rg$ and varying the maximum viscous frequency from 2 Hz to 30 Hz. The resulting lag spectra, illustrated in Fig. \ref{fig:vfv}, show that the zero-crossing points of the soft lags remain consistent at approximately 50 Hz, while the hard lags shift to higher frequencies, accompanied by a decrease in the soft lags. 

\begin{figure*}[!p]
   \centering
   \subfigure[\label{fig:vgeo}]{\includegraphics[width=0.4\linewidth]{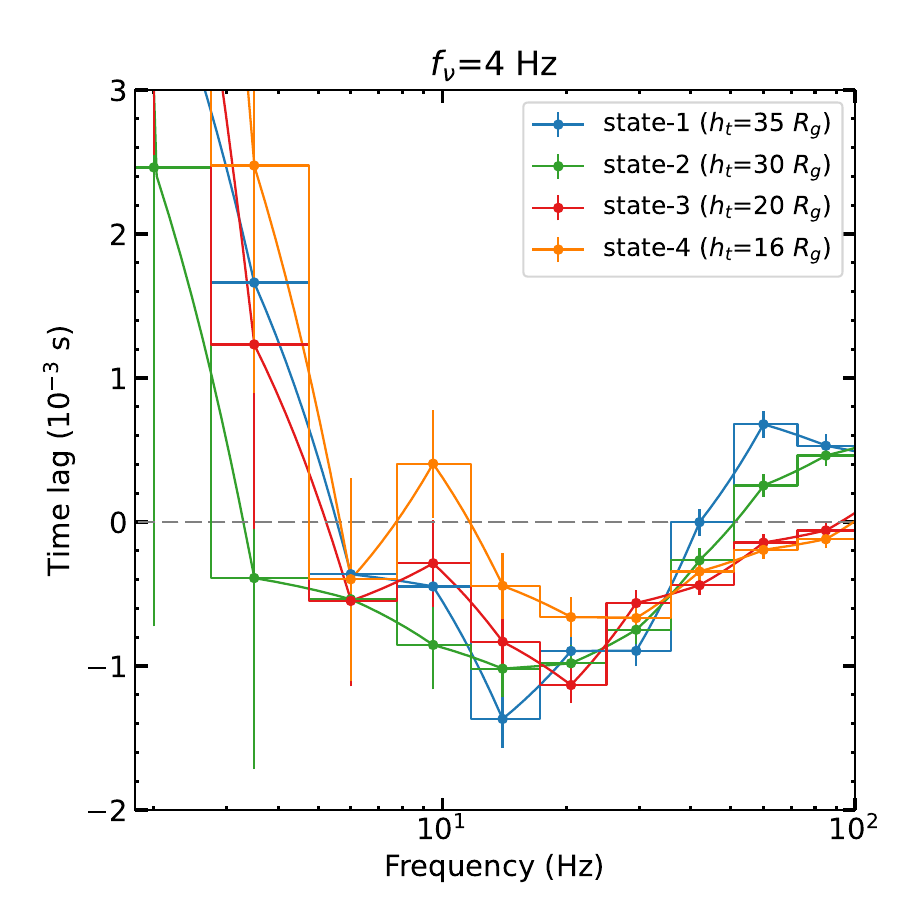}}\hspace{20pt}
   \subfigure[\label{fig:vfv}]{\includegraphics[width=0.4\linewidth]{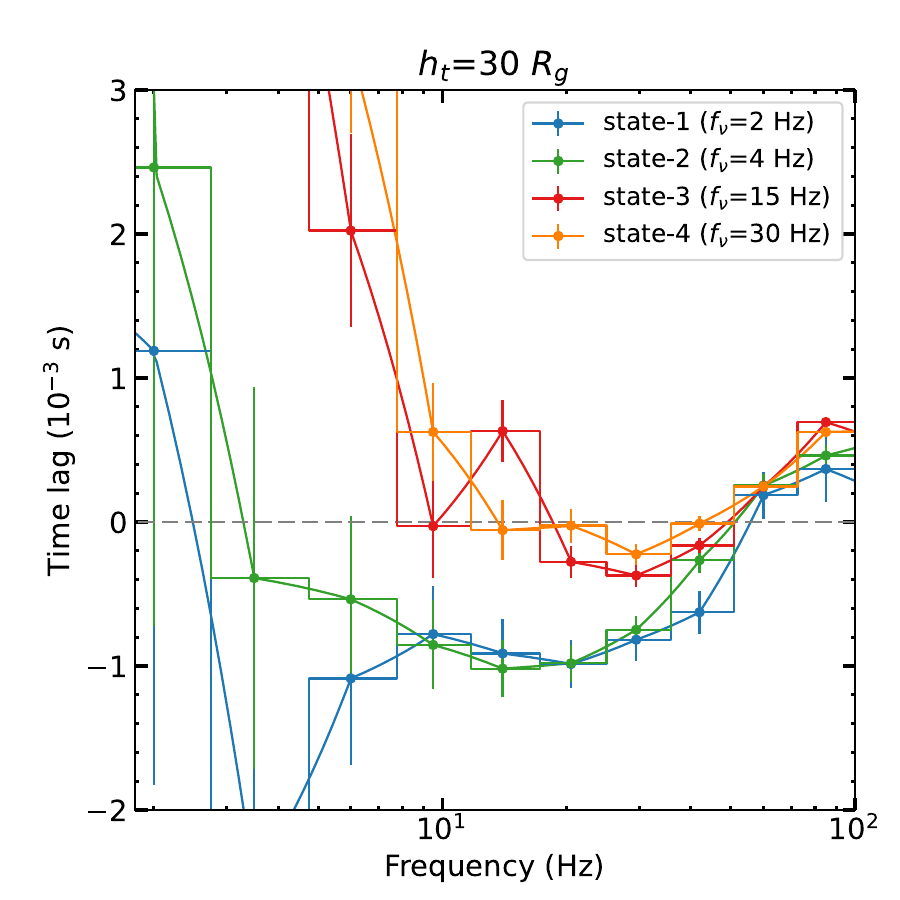}}
  \caption{(a) When the maximum viscous frequency is fixed at 4 Hz and the height of the corona decreases from \(35 \Rg\) to \(16 \Rg\) (corresponding to the evolution from state-1 to state-4), the soft lags slightly decrease, and the zero-crossing points monotonically shift to higher frequencies. (b) When the height of the corona is fixed at \(30 \Rg\) and the maximum viscous frequency increases from 4 Hz to 30 Hz (corresponding to the evolution from state-1 to state-4), the hard lags shift to higher frequencies, and the amplitudes of the soft lags decrease. However, the zero-crossing points remain unchanged. The parameters in each state are listed in the legend.}\label{fig:vgeo-vfv}
\end{figure*}

In Fig. \ref{fig:vgeo-vfv}, we explore the effects of varying the coronal geometry and maximum viscous frequency separately. In Fig. \ref{fig:vgeo}, the maximum viscous frequency is held constant at 4 Hz, and the height of the corona varies from $35~R_g$ to $16~R_g$. We see that, as the coronal height is reduced, the amplitude of the soft lag decreases and the zero-crossing point shifts to higher frequencies (from $\sim 40 Hz$ to $\sim 100$ Hz). In Fig. \ref{fig:vfv}, the coronal height is instead held constant at $30~R_g$ and the maximum viscous frequency is varied from 2 Hz to 30 Hz. As the maximum viscous frequency increases, the hard lag shifts to higher frequencies, and the amplitude of the soft lag decreases, but the zero-crossing point remains constant at $\sim 50$ Hz.

The independent changes in geometry and viscosity produce distinct lag patterns, indicating that the geometry and the viscosity play different roles in the evolution of lags. We will further discuss this in Section \ref{sec:discuss}.

% \hspace{-8cm}
% \begin{table}[h]
% \renewcommand{\arraystretch}{1.2}
%  % \setlength{\tabcolsep}{-6pt}
% \caption{The list of maximum viscous frequencies and height in each figure.}\label{tab:para}
%  \centering
% \begin{tabular}{@{}cccccccccc@{}}
% \toprule
% &  & & & state-1 & state-2 & state-3 & state-4 &       &      \\ \midrule
% \multirow{2}{*}{\begin{tabular}[c]{@{}c@{}}simulated evolution\\ (Fig. \ref{fig:lag-spectra})\end{tabular}} &  & &  $f_\text{max}$ & 2  & 4 & 15  & 30      & $\text{Hz}$ & \\
%        & &    & $h_t$          & 35      & 30      & 20      & 16      & $R_g$   &    \\ \midrule

% \multirow{2}{*}{\begin{tabular}[c]{@{}c@{}}Fix $f_\text{max}$\\ (Fig. \ref{fig:vgeo})\end{tabular}} &   & & $f_\text{max}$ & 4 & 4 & 4   & 4 & $\text{Hz}$& \\
%    & && $h_t$   & 35& 30  & 20   & 16    & $R_g$  &     \\ \midrule

% \multirow{2}{*}{\begin{tabular}[c]{@{}c@{}}Fix $h_t$\\ (Fig. \ref{fig:vfv})\end{tabular}} &   & &$f_\text{max}$ &  2  & 4    & 15  & 30& $\text{Hz}$ &\\
%   & & & $h_t$          & 30      & 30   & 30      & 30      & $R_g$   &    \\ \bottomrule
% \end{tabular}
% \end{table}
\section{Discussion \label{sec:discuss}}

\par There are several factors that may influence the inter-band lags. For instance, viscous propagation from the disc to the corona can raise the temperature of the corona and lead to fluctuations in coronal temperature. These fluctuations may contribute to the occurrence of hard lags. In addition, partially Comptonized photons can heat the disc, resulting in variable thermal emission, which can lead to soft lags \citep{uttleyLargeComplexXray2024}. 
Despite these possibilities, this study ignores all heating effects and temperature fluctuations for the sake of simplicity. It's necessary to note that ignoring the heating effects may require a larger geometrical size of the corona and a larger reflection region to reproduce the observed lag. Instead, we concentrate on the delays associated with the propagating fluctuations and the light travel times involved in the inter-band lags (Sect. \ref{sec:the effect}). Furthermore, we will examine how the evolution of the corona affects the frequency-dependent time lags (Sect. \ref{sec:the evolution}). Besides, we will show that predictions from 
%both the vertically extended and radially extended corona models 
both a jet base corona model and a spherical corona model 
are consistent with observational data (Sect. \ref{sec:the_geometry}). This suggests that simulating X-ray polarization will be essential for further distinguishing the geometry of the corona in future research \citep{krawczynskiPolarizedXraysConstrain2022, kravtsovOpticalPolarizationSignatures2022, podgornySpectralPolarizationProperties2022}.

\par In typical timing spectral analysis, the energy-dependent lag spectrum is a good probe to understand the physical processes behind the lags at different energies in a specific frequency band \citep{arevaloSpectraltimingEvidenceVery2006,cassatellaAccretionFlowDiagnostics2012,demarcoTRACINGREVERBERATIONLAG2015}. The energy-dependent lag spectrum of MAXI J1820+070 shows that in the low frequency of 0.1-1 Hz, the lags increase as the energies increase in the energy band of 0.5-1 keV \citep{karaCoronaContractsBlackhole2019}. This may originate from the propagation of fluctuations from a low energy area to high energy area, causing propagating lags. We tried to reproduce the energy-dependent lag spectrum from our simulation. We found that the simulated lag matches the the trend of the observed spectrum in the energy band below 1 keV, however, it breaks to a constant in the energy band over 1 keV. It's necessary to note that the disk component vanishes over 1 keV, as shown in Fig. \ref{fig:Espec}, and propagation of fluctuation in the corona is not considered in this work. Thus, after the disk component vanishes, the fluctuation can not propagate to the higher energy area, which may explain why the lags do not increase over 1 keV. We will implement the propagation of fluctuation in the corona in future work.

\subsection{The roles of the propagating fluctuation and the light traveling in the inter-band lag}
\label{sec:the effect}

\par
\begin{figure*}[htbp]
    \centering
    \includegraphics[width=0.75\linewidth]{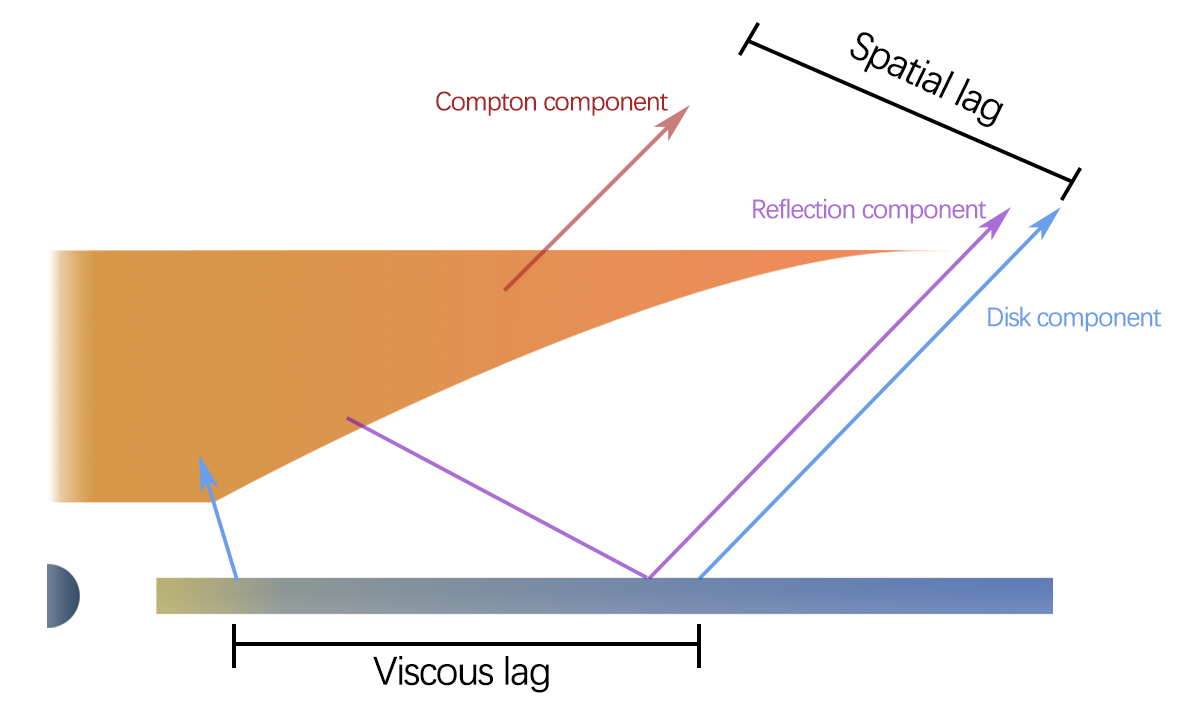}
    \caption{The colors on the disc serve as a visual representation of the variability speed across the disc, where a deeper yellow intensity indicates a higher viscous frequency. The corona obstructs the partial accretion disc, and due to the emission profile on the disc, the disc component originates from the outer region, while the Comptonized photons come from the inner region. This results in fluctuations of the Compton and reflection components occurring on the timescale of the inner region. The propagation lag between the outer and inner regions, which occurs over several seconds, is greater than the light-travel delay that occurs over milliseconds. As a result, the propagation lag predominates in the low-frequency band, leading to the observed low-frequency hard lag. Additionally, this propagation lag diminishes at higher frequencies, where spatial lags take precedence. The travel time for soft thermal photons reaching the corona is several dozen $\Rg/c$, which is shorter than the travel time for hard Comptonized photons interacting with the disc, measured in hundreds of $\Rg/c$. Consequently, at high frequencies, this longer travel delay for hard photons results in the soft lag.}
    \label{fig:lag_mechanism}
\end{figure*}

 In Sect. \ref{sec:result}, Fig. \ref{fig:lag-spectra-single} shows 
. We briefly analyze the viscously propagating lag, hereafter propagation lag, caused by the delay in propagating fluctuation, and the spatial lag caused by the delay in light travel. Here, we describe how the spatial lags and propagation lags result in the lag spectra with Fig. \ref{fig:lag_mechanism}.

\par As summarized in Section \ref{sec:result}, the inclination angle is set to \( i = 60^{\circ} \). In this scenario, the inner region of the accretion disc is obscured by the corona, meaning that the observed disc component originates from the outer region, specifically from approximately \( r > 130 \Rg \), based on the assumed geometry. Meanwhile, the seed photons for Comptonization primarily come from the inner region of the disc due to its emission profile. 
The position of these seed photons suggests that the Compton and reflection components fluctuate according to the variability from the inner region. Considering the variability of these spectral components, we anticipate that the propagation lag between the inner (high-energy) and outer (low-energy) regions, measured on a timescale of seconds, exceeds the spatial lag, which occurs on a timescale of milliseconds. Consequently, the propagation lag is more significant than the spatial lag in the low-frequency band. This observation accounts for the low-frequency hard lag illustrated in Fig. \ref{fig:lag-spectra-single}.

It is important to note that the propagation lag, which occurs over longer timescales, diminishes as the Fourier frequency increases. Therefore, in the high-frequency band, the spatial lag becomes dominant over the propagation lag. Additionally, the travel delay for low-energy thermal photons reaching the corona is several tens of \( \Rg/c \), while the travel delay for high-energy Comptonized photons returning to the disc is on the order of hundreds of \( \Rg/c \). Thus, at high frequencies, the latter delay is more significant, resulting in the high-frequency soft lag shown in Fig. \ref{fig:lag-spectra-single}.

\subsection{Observational test: explanation of the evolution of the lags in MAXI J1820+070}
\label{sec:the evolution}

\par The observed frequency-dependent lag shows a trend that the zero-crossing points shift toward higher frequency and the amplitudes of soft lags decrease, indicating the evolution of the inner accretion flow \citep{karaCoronaContractsBlackhole2019,demarcoInnerFlowGeometry2021}. These observational results can be successfully reproduced in the simulations in this work, by changing both the geometry of the corona and the maximum viscous frequency (Fig. \ref{fig:lag-spectra}). In this section, we will show that the evolution of the lag spectrum can be attributed to the change of both the geometry and the maximum viscous frequency, as is shown in Fig. (\ref{fig:vgeo-vfv}). 

% \begin{figure*}[h]
%    \centering
%    \subfigure[\label{fig:vgeo}]{\includegraphics[width=0.45\linewidth]{4geo_modified.pdf}}\hspace{20pt}
%    \subfigure[\label{fig:vfv}]{\includegraphics[width=0.45\linewidth]{4vfv_modified.pdf}}
%   \caption{(a) When the maximum viscous frequency is fixed at 4 Hz and the height of the corona decreases from \(35 \Rg\) to \(16 \Rg\) (corresponding to the evolution from state-1 to state-4), the soft lags slightly decrease, and the zero-crossing points monotonically shift to higher frequencies. (b) When the height of the corona is fixed at \(30 \Rg\) and the maximum viscous frequency increases from 4 Hz to 30 Hz (corresponding to the evolution from state-1 to state-4), the hard lags shift to higher frequencies, and the amplitudes of the soft lags decrease. However, the zero-crossing points remain unchanged. The parameters in each state are listed in Tab. \ref{tab:para}.\textbf{}}
% \end{figure*}

% \par We investigate the effects of geometrical changes by maintaining the maximum viscous frequency at 4 Hz while reducing the height of the corona from \(35 ~\Rg\) to \(16 ~\Rg\). As shown in Fig. \ref{fig:vgeo}, when the frequency exceeds 4 Hz, the amplitude of the lag decreases, and the zero-crossing point for the soft lag shifts monotonically to higher frequencies. 

\par
\begin{figure*}[htbp]
   \centering
   \subfigure[\label{fig:sch_vgeo}]{\includegraphics[width=0.8\linewidth]{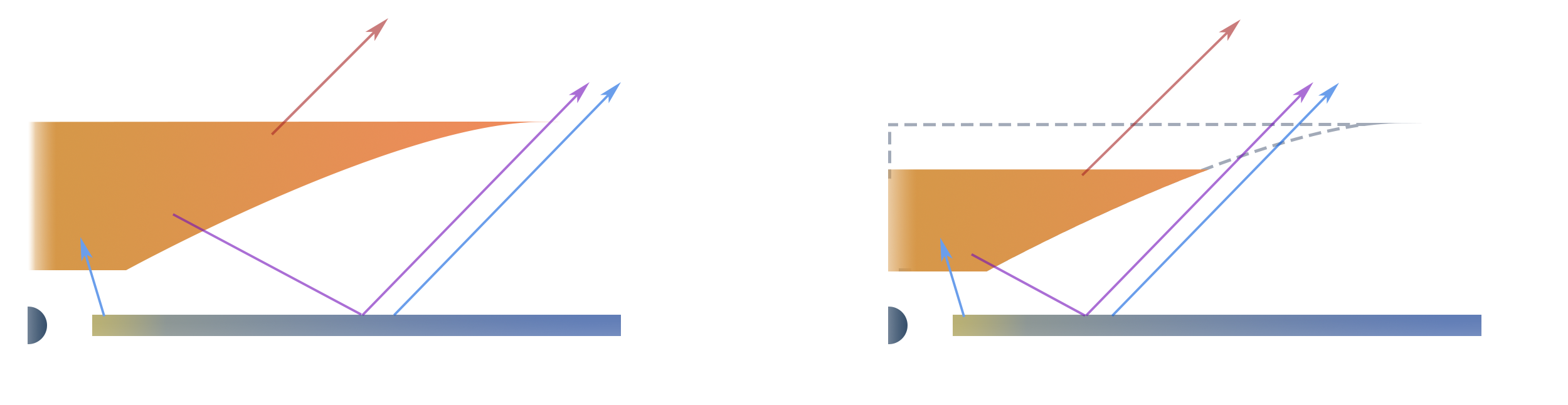}}\hspace{20pt}
   \subfigure[\label{fig:sch_vfv}]{\includegraphics[width=0.8\linewidth]{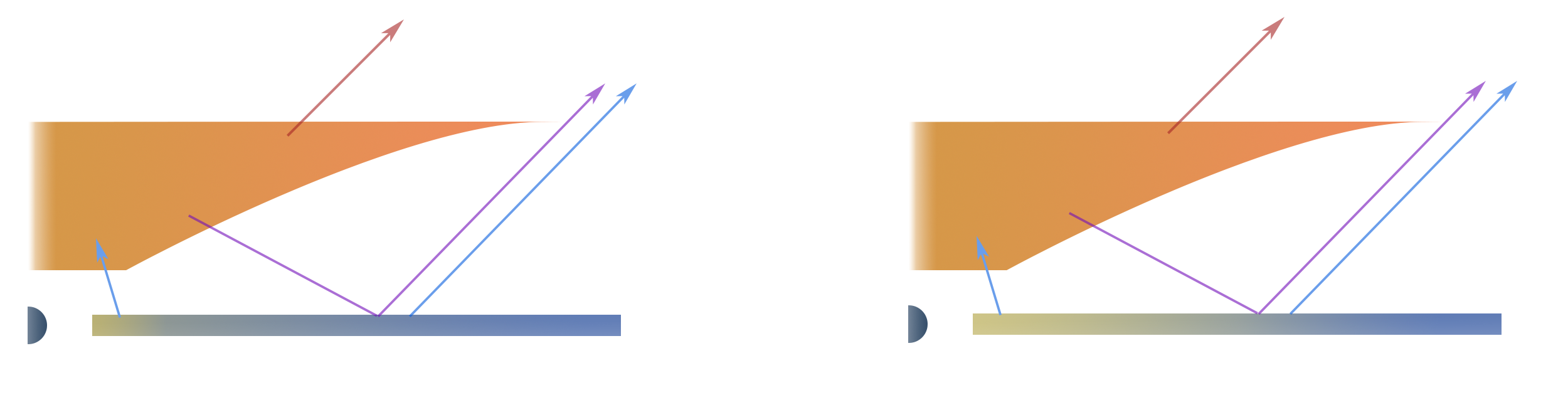}}
   \caption{The colors on the disc serve as a visual representation of the variability speed across the disc. A deeper yellow intensity indicates a higher viscous frequency. (a) When the height of the corona decreases, the spatial lag of the Comptonized photons striking the disc also decreases, leading to a reduction in high-frequency soft lags and a shift in the zero-crossing points for these soft lags. (b) As the maximum viscous frequency increases, the timescales of variability in each region become shorter, causing the propagation lags to dominate over the spatial lag at higher frequencies. This results in a transition from low-frequency lags to shorter timescales. Additionally, since the geometry of the corona remains unchanged, the zero-crossing points also remain constant.}
   \label{fig:lagevomechanic}
\end{figure*}

 The changes in geometry play a role in the evolution of the lag at high frequency. As demonstrated in Fig. \ref{fig:vgeo}, the zero-crossing point shifts toward higher frequencies as the height of the corona decreases. As the separation between the corona and the disc decreases (see Fig. \ref{fig:sch_vgeo}), the spatial lag for the Comptonized photons striking the disc also diminishes, resulting in a reduction of the soft lag. Furthermore, a shorter soft lag causes its zero-crossing point to shift toward higher frequencies \citep{mizumotoXrayShorttimeLags2018, demarcoInnerFlowGeometry2021, yuSpectraltimingStudyInner2023}. Thus, the decrease in spatial lag due to the geometrical change of the corona accounts for the reduced amplitude of the lag and the shift of the zero-crossing points to higher frequencies, as illustrated in Fig. \ref{fig:lag-spectra}.

% The evolution of the lag may come from the evolution of both viscous and spatial lag.
% \par The observed and simulated lag spectra show a trend to vary in a shorter timescale, influencing the zero points to shift towards higher frequency and the amplitude of soft lags to decrease. The evolution of the lag may come from the evolution of both viscous and spatial lag. To figure out the individual influence of viscous and spatial lag, we ran two simulations with two isolated parameters of the height of the corona and the maximum viscous frequency, also MVF. 

% Upon keeping the MVF constant while decreasing the corona's height, the lag's amplitude was found to decrease the second zero points shift to high frequency monotonically (see Fig. \ref{fig:vgeo}). 
% aaa

% In addition, when the height of the corona remains fixed, and the MVF increases, the positions of the first zero points is seen to transition from low to high frequency, while the second zero points are clearly remain unaffected by these alterations. Simultaneously, there is a notable decrease of the lag's amplitudes (see Fig. \ref{fig:vfv}).

% v
\par The changes in viscous frequency play a role in the evolution in the "low-frequency" band (around several hertz). Illustrating in Fig. \ref{fig:vfv}, the low-frequency lags shift towards high-frequency as the maximum viscous frequency increases. According to the formula for the viscous frequency at each radius, \( f_{\mathrm{visc}}(r_n) = f_{\max}\left(r_n/r_o\right)^{-3/2} \), discussed in Sect. \ref{sec:fluctuations}, when the maximum viscous frequency increases, the characteristic timescale at each radius increases, as shown in Fig. \ref{fig:sch_vfv}. This results in faster time variability for all emissions, causing propagation lags to dominate over spatial lags at higher frequencies. Consequently, the low-frequency lags shift to shorter timescales in Fig. \ref{fig:lag-spectra}. Furthermore, since the geometry of the corona remains unchanged, the intrinsic amplitude of the reverberation lags stays constant, which explains the unchanged zero-crossing points in Fig. \ref{fig:vfv}.

% \par The decrease of the height of the corona and the increase of maximum viscous frequency are enough to explain the evolution pattern of the lag spectra. But the physical nature of MVF's increase is unknown. According to the numerical solution of $\alpha$-disc for viscous frequency $f_{\text{visc}} \sim 3 \times 10^{-6} \alpha^{-4 / 5} \dot{M}_{16}^{3 / 10} m_{1}^{-1 / 4} R_{10}^{-5 / 4} \mathrm{~Hz}.$ \citep{frankAccretionPowerAstrophysics2002}, the maximum viscous frequency can reasonably increase by increasing $\alpha$ due to some physical processes, like an enhance of magnetic field \citep{youObservationsBlackHole2023}. 

% \subsection{Radially-extended corona}
\subsection{Alternative spherical corona case}
\label{sec:the_geometry}

\par
\begin{figure*}[!p]
   \centering
   \subfigure[\label{fig:adaf-geo}]{\includegraphics[width=0.45\linewidth]{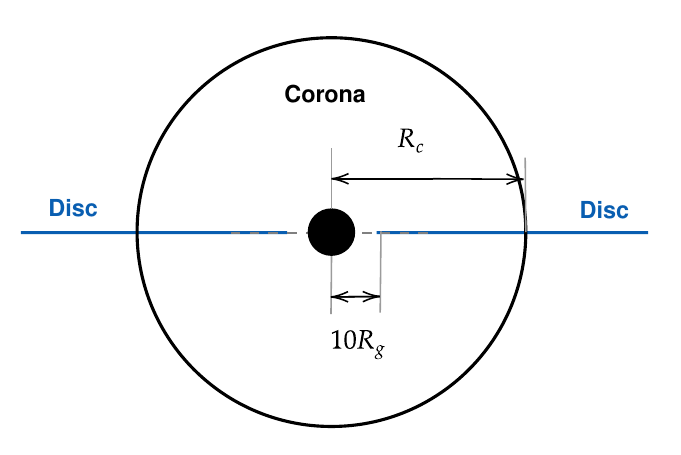}}\hspace{20pt}
       \subfigure[\label{fig:adaf}]{\includegraphics[width=0.4\linewidth]{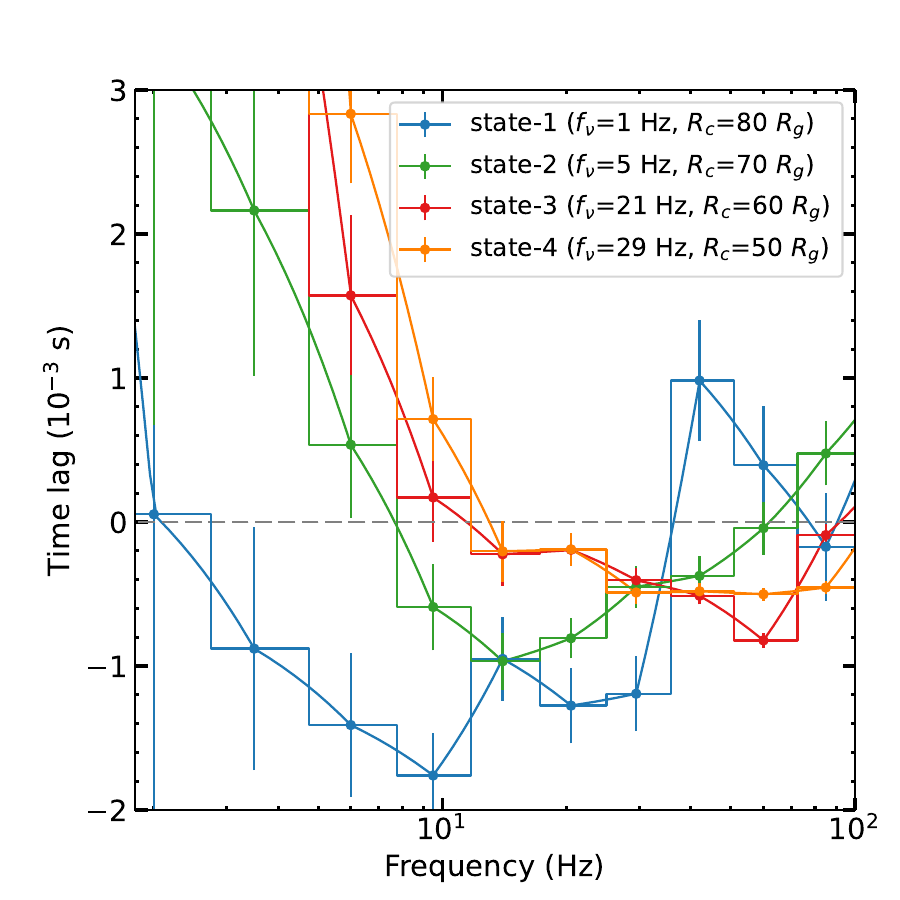}}
  \caption{(a) We present a spherical corona as the simplest case of an extended corona. The accretion disc extends into the corona with a small truncated radius $10\Rg$. The radius of the corona decreases from $80\Rg$ (state-1) to $50\Rg$ (state-4), and the maximum viscous frequency increases from 1 Hz to 29 Hz for each state.  
  (b) The Lag spectrum of a spherical corona. The spectrum shows a high-frequency soft lag and low-frequency hard lag as well as the evolution pattern of lags, which are consistent with the observations.}
\end{figure*}

 In our main result, we focus on a corona that is wider than it is tall but does not exclude the possibility of another extended model. To assess the feasibility of other extended models, we performed a simulation using the simplest extended geometry that incorporates a spherical corona centered around a black hole alongside an accretion disc that extends into the corona. In this configuration, the disc's expansion minimizes the influence of the disk component, allowing for a more manageable evaluation. We set the radius of the spherical corona at \(R_c\), while the extended disc is truncated at a smaller radius of \(10 R_g\) (as shown in Fig. \ref{fig:adaf-geo}). In the same way as the simulation in the main content, we simulated 4 states with increasing radii of the corona and decreasing maximum viscous frequencies. The radius of the corona decreases from $80\Rg$ (state-1) to $50\Rg$ (state-4), and the maximum viscous frequency increases from 1 Hz to 29 Hz for each state. The derived frequency-dependent lag predicted from the simulation is plotted in Fig. \ref{fig:adaf} in which the low-frequency hard lag and the high-frequency soft lag, as well as the evolution pattern of lags, are both observed. This suggests that the model with a spherical corona could also quantitatively explain the observed data, indicating that the corona's geometry remains unresolved in this analysis.

\par Further observations and simulations are needed to distinguish the explicit geometry of the corona clearly. X-ray polarization detection is a highly effective method for probing this geometry. Recent findings from the Imaging X-ray Polarimetry Explorer (IXPE) indicate that the corona is extended perpendicular to the resolved radio jet for several X-ray binaries \citep{krawczynskiPolarizedXraysConstrain2022,ingramTrackingXRayPolarization2024}. The spherical corona model, which can explain the timing results as well as any other geometry, is thus ruled out by IXPE results, because it predicts the coronal emission to be unpolarized. The jet base model, in contrast, can be consistent with IXPE results if the parameters are chosen to ensure the corona is wider than it is tall, as they are in this paper. The two mechanisms we explore in this paper to vary the soft lag amplitude will give rise to very different polarization signals. The polarization should stay constant as the maximum viscous frequency increases, whereas the polarization degree is expected to increase as the coronal height reduces, thus making the corona even more vertically extended. Dense IXPE monitoring of future X-ray binary outbursts will therefore provide a powerful diagnostic. IXPE monitoring of Swift J1727.8-1613 demonstrated that the aspect ratio of the corona remains almost constant during the hard intermediate state, even though the soft lag amplitude increases dramatically \citep{ingramTrackingXRayPolarization2024}. However, the epoch of the hard state in which the soft lag amplitude reduces was already over before the first IXPE observation took place.
% have provided valuable constraints on the geometry of X-ray sources in certain BHXRBs.
With the upcoming launch of the Enhanced X-ray Timing and Polarimetry mission (eXTP), additional polarimetry observations will become available, necessitating further simulations of X-ray polarization to refine our understanding of the corona's geometry. Therefore, we plan to conduct polarimetry simulations for various extended corona models in our future research.

%This means that the simulation of the X-ray polarization is also required to distinguish the corona geometry in future work Cyg X-1 in the hard state and for Swift J1727.8–1613 in the HIMS favour a more horizontally extended corona (Kraw

\section{Acknowledgment}
    We thank Stefano Rapisarda for his help in modeling the propagating fluctuation, and Phil Uttley for his suggestion regarding the implication of the simulation. We also thank Wei Yu, Liang Zhang, Wenda Zhang, and Honghui Liu for their helpful discussion/comments. We also thank Yongkang Zhang for his help in preparing the schematic. B.Y. is supported by Natural Science Foundation of China (NSFC) grants 12322307, 12361131579, and 12273026; by the National Program on Key Research and Development Project 2021YFA0718500; by Xiaomi Foundation / Xiaomi Young Talents Program. FYW is supported by the Natural Science Foundation of China (NSFC) grants 12494575 and 12273009. A.I. acknowledges support from the Royal Society.
\newpage
    
    \bibliographystyle{aasjournal}
    \bibliography{x-ray.bib}

\begin{thebibliography}{}
\expandafter\ifx\csname natexlab\endcsname\relax\def\natexlab#1{#1}\fi
\providecommand{\url}[1]{\href{#1}{#1}}
\providecommand{\dodoi}[1]{doi:~\href{http://doi.org/#1}{\nolinkurl{#1}}}
\providecommand{\doeprint}[1]{\href{http://ascl.net/#1}{\nolinkurl{http://ascl.net/#1}}}
\providecommand{\doarXiv}[1]{\href{https://arxiv.org/abs/#1}{\nolinkurl{https://arxiv.org/abs/#1}}}

\bibitem[{Alston {et~al.}(2020)Alston, Fabian, Kara, Parker, Dovciak, Pinto, Jiang, Middleton, Miniutti, Walton, Wilkins, Buisson, {Caballero-Garcia}, Cackett, De~Marco, Gallo, Lohfink, Reynolds, Uttley, Young, \& Zogbhi}]{alstonDynamicBlackHole2020}
Alston, W.~N., Fabian, A.~C., Kara, E., {et~al.} 2020, Nature Astronomy, 4, 597, \dodoi{10.1038/s41550-019-1002-x}

\bibitem[{Ar{\'e}valo {et~al.}(2006)Ar{\'e}valo, Papadakis, Uttley, McHardy, \& Brinkmann}]{arevaloSpectraltimingEvidenceVery2006}
Ar{\'e}valo, P., Papadakis, I.~E., Uttley, P., McHardy, I.~M., \& Brinkmann, W. 2006, Monthly Notices of the Royal Astronomical Society, 372, 401, \dodoi{10.1111/j.1365-2966.2006.10871.x}

\bibitem[{Ar{\'e}valo \& Uttley(2006)}]{arevaloInvestigatingFluctuatingaccretionModel2006}
Ar{\'e}valo, P., \& Uttley, P. 2006, Monthly Notices of the Royal Astronomical Society, 367, 801, \dodoi{10.1111/j.1365-2966.2006.09989.x}

\bibitem[{Belloni(2010)}]{belloniStatesTransitionsBlack2010}
Belloni, T. 2010, in The {{Jet Paradigm}}, ed. T.~Belloni, Vol. 794 (Berlin, Heidelberg: Springer Berlin Heidelberg), 53--84, \dodoi{10.1007/978-3-540-76937-8_3}

\bibitem[{Bendat \& Piersol(2011)}]{bendatRandomDataAnalysis2011}
Bendat, J.~S., \& Piersol, A.~G. 2011, Random {{Data}}: {{Analysis}} and {{Measurement Procedures}} (John Wiley \& Sons)

\bibitem[{Bhattacharyya \& Strohmayer(2007)}]{bhattacharyyaEvidenceBroadRelativistic2007}
Bhattacharyya, S., \& Strohmayer, T.~E. 2007, The Astrophysical Journal, 664, L103, \dodoi{10.1086/520844}

\bibitem[{Bollemeijer {et~al.}(2024)Bollemeijer, Uttley, Basak, Ingram, {van~den~Eijnden}, Alabarta, Altamirano, Arzoumanian, Buisson, Fabian, Ferrara, Gendreau, Homan, Kara, Markwardt, Remillard, Sanna, Steiner, Tombesi, Wang, Wang, \& Zoghbi}]{bollemeijerEvidenceDynamicCorona2024}
Bollemeijer, N., Uttley, P., Basak, A., {et~al.} 2024, Monthly Notices of the Royal Astronomical Society, 528, 558, \dodoi{10.1093/mnras/stad3912}

\bibitem[{Bu {et~al.}(2021)Bu, Zhang, Santangelo, Belloni, Zhang, Qu, Tao, Huang, Ma, Li, Zhang, Chen, {collaboration}, Cai, Cao, Chang, Chen, Chen, Chen, Cui, Du, Gao, Gao, Ge, Gu, Guan, Guo, Han, Huo, Jia, Jiang, Jin, Kong, Li, Li, Li, Li, Li, Li, Li, Li, Li, Liang, Liao, Liu, Liu, Liu, Liu, Lu, Lu, Luo, Luo, Ma, Meng, Nang, Nie, Ou, Sai, Song, Song, Sun, Tan, Tuo, Wang, Wang, Wang, Wang, Wang, Wen, Wu, Wu, Wu, Xiao, Xiao, Xiong, Xu, Yang, Yang, Yi, Yin, You, Zhang, Zhang, Zhang, Zhang, Zhang, Zhang, Zhang, Zhang, Zhao, Zhao, Zheng, \& Zhou}]{buBroadbandVariabilityStudy2021}
Bu, Q.~C., Zhang, S.~N., Santangelo, A., {et~al.} 2021, The Astrophysical Journal, 919, 92, \dodoi{10.3847/1538-4357/ac11f5}

\bibitem[{Buisson {et~al.}(2019)Buisson, Fabian, Barret, F{\"u}rst, Gandhi, Garc{\'i}a, Kara, Madsen, Miller, Parker, Shaw, Tomsick, \& Walton}]{buissonMAXIJ1820+070NuSTAR2019}
Buisson, D. J.~K., Fabian, A.~C., Barret, D., {et~al.} 2019, Monthly Notices of the Royal Astronomical Society, 490, 1350, \dodoi{10.1093/mnras/stz2681}

\bibitem[{Cassatella {et~al.}(2012)Cassatella, Uttley, \& Maccarone}]{cassatellaAccretionFlowDiagnostics2012}
Cassatella, P., Uttley, P., \& Maccarone, T.~J. 2012, Monthly Notices of the Royal Astronomical Society, 427, 2985, \dodoi{10.1111/j.1365-2966.2012.22021.x}

\bibitem[{Chainakun \& Young(2015)}]{chainakunSimultaneousSpectralReverberation2015}
Chainakun, P., \& Young, A.~J. 2015, Monthly Notices of the Royal Astronomical Society, 452, 333, \dodoi{10.1093/mnras/stv1333}

\bibitem[{Chand {et~al.}(2022)Chand, Dewangan, Thakur, Tripathi, \& Agrawal}]{chandAstroSatViewNewly2022}
Chand, S., Dewangan, G.~C., Thakur, P., Tripathi, P., \& Agrawal, V.~K. 2022, The Astrophysical Journal, 933, 69, \dodoi{10.3847/1538-4357/ac7154}

\bibitem[{Dauser {et~al.}(2016)Dauser, Garcia, Walton, Eikmann, Kallman, McClintock, \& Wilms}]{dauserNormalizingRelativisticModel2016}
Dauser, T., Garcia, J., Walton, D.~J., {et~al.} 2016, Astronomy \& Astrophysics, 590, A76, \dodoi{10.1051/0004-6361/201628135}

\bibitem[{Dauser {et~al.}(2013)Dauser, Garcia, Wilms, B{\"o}ck, Brenneman, Falanga, Fukumura, \& Reynolds}]{dauserIrradiationAccretionDisc2013}
Dauser, T., Garcia, J., Wilms, J., {et~al.} 2013, Monthly Notices of the Royal Astronomical Society, 430, 1694, \dodoi{10.1093/mnras/sts710}

\bibitem[{Dauser {et~al.}(2022)Dauser, Garc{\'i}a, Joyce, Licklederer, Connors, Ingram, Reynolds, \& Wilms}]{dauserEffectReturningRadiation2022}
Dauser, T., Garc{\'i}a, J.~A., Joyce, A., {et~al.} 2022, Monthly Notices of the Royal Astronomical Society, 514, 3965, \dodoi{10.1093/mnras/stac1593}

\bibitem[{De~Marco {et~al.}(2022)De~Marco, Motta, \& Belloni}]{demarcoProbingBlackHoleAccretion2022}
De~Marco, B., Motta, S.~E., \& Belloni, T.~M. 2022, in Handbook of {{X-ray}} and {{Gamma-ray Astrophysics}}, ed. C.~Bambi \& A.~Santangelo (Singapore: Springer Nature), 1--41, \dodoi{10.1007/978-981-16-4544-0_129-1}

\bibitem[{De~Marco \& Ponti(2019)}]{demarcoObservationsXrayReverberation2019}
De~Marco, B., \& Ponti, G. 2019, Astronomische Nachrichten, 340, 290, \dodoi{10.1002/asna.201913612}

\bibitem[{De~Marco {et~al.}(2015)De~Marco, Ponti, {Mu{\~n}oz-Darias}, \& Nandra}]{demarcoTRACINGREVERBERATIONLAG2015}
De~Marco, B., Ponti, G., {Mu{\~n}oz-Darias}, T., \& Nandra, K. 2015, The Astrophysical Journal, 814, 50, \dodoi{10.1088/0004-637X/814/1/50}

\bibitem[{De~Marco {et~al.}(2021)De~Marco, Zdziarski, Ponti, Migliori, Belloni, Otero, Dzie{\l}ak, \& Lai}]{demarcoInnerFlowGeometry2021}
De~Marco, B., Zdziarski, A.~A., Ponti, G., {et~al.} 2021, Astronomy \& Astrophysics, 654, A14, \dodoi{10.1051/0004-6361/202140567}

\bibitem[{De~Marco {et~al.}(2017)De~Marco, Ponti, Petrucci, Clavel, Corbel, Belmont, Chakravorty, Coriat, Drappeau, Ferreira, Henri, Malzac, Rodriguez, Tomsick, Ursini, \& Zdziarski}]{demarcoEvolutionReverberationLag2017}
De~Marco, B., Ponti, G., Petrucci, P.~O., {et~al.} 2017, Monthly Notices of the Royal Astronomical Society, 471, 1475, \dodoi{10.1093/mnras/stx1649}

\bibitem[{Done {et~al.}(2007)Done, Gierli{\'n}ski, \& Kubota}]{doneModellingBehaviourAccretion2007}
Done, C., Gierli{\'n}ski, M., \& Kubota, A. 2007, The Astronomy and Astrophysics Review, 15, 1, \dodoi{10.1007/s00159-007-0006-1}

\bibitem[{Dovciak(2004)}]{dovciakRadiationAccretionDiscs2004}
Dovciak, M. 2004, Phd Thesis, 27, 238, \dodoi{10.48550/arXiv.astro-ph/0411605}

\bibitem[{Dubus {et~al.}(2001)Dubus, Hameury, \& Lasota}]{dubusDiscInstabilityModel2001}
Dubus, G., Hameury, J.-M., \& Lasota, J.-P. 2001, Astronomy \& Astrophysics, 373, 251, \dodoi{10.1051/0004-6361:20010632}

\bibitem[{Emmanoulopoulos {et~al.}(2014)Emmanoulopoulos, Papadakis, Dov{\v c}iak, \& McHardy}]{emmanoulopoulosGeneralRelativisticModelling2014}
Emmanoulopoulos, D., Papadakis, I.~E., Dov{\v c}iak, M., \& McHardy, I.~M. 2014, Monthly Notices of the Royal Astronomical Society, 439, 3931, \dodoi{10.1093/mnras/stu249}

\bibitem[{Esin {et~al.}(1997)Esin, McClintock, \& Narayan}]{esinAdvectionDominatedAccretionSpectral1997}
Esin, A.~A., McClintock, J.~E., \& Narayan, R. 1997, The Astrophysical Journal, 489, 865, \dodoi{10.1086/304829}

\bibitem[{Gao {et~al.}(2023)Gao, Yan, \& Yu}]{gaoLowfrequencyQuasiperiodicOscillation2023}
Gao, C., Yan, Z., \& Yu, W. 2023, Monthly Notices of the Royal Astronomical Society, 520, 5544, \dodoi{10.1093/mnras/stad434}

\bibitem[{Garc{\'i}a {et~al.}(2013)Garc{\'i}a, Dauser, Reynolds, Kallman, McClintock, Wilms, \& Eikmann}]{garciaXRAYREFLECTEDSPECTRA2013}
Garc{\'i}a, J., Dauser, T., Reynolds, C.~S., {et~al.} 2013, The Astrophysical Journal, 768, 146, \dodoi{10.1088/0004-637X/768/2/146}

\bibitem[{George \& Fabian(1991)}]{georgeXrayReflectionCold1991}
George, I.~M., \& Fabian, A.~C. 1991, Monthly Notices of the Royal Astronomical Society, 249, 352, \dodoi{10.1093/mnras/249.2.352}

\bibitem[{Gorecki \& Wilczewski(1984)}]{goreckiStudyComptonizationRadiation1984}
Gorecki, A., \& Wilczewski, W. 1984, Acta Astronomica, 34, 141

\bibitem[{Haardt \& Maraschi(1991)}]{haardtTwoPhaseModelXRay1991}
Haardt, F., \& Maraschi, L. 1991, The Astrophysical Journal, 380, L51, \dodoi{10.1086/186171}

\bibitem[{Homan \& Belloni(2005)}]{homanEvolutionBlackHole2005}
Homan, J., \& Belloni, T. 2005, in Astrophysics and {{Space Science}}: {{From X-Ray Binaries}} to {{Quasars}}: {{Black Holes}} on All {{Mass Scales}}, ed. T.~J. Maccarone, R.~P. Fender, \& L.~C. Ho (Dordrecht: Springer Netherlands), 107--117, \dodoi{10.1007/1-4020-4085-7_13}

\bibitem[{Huang {et~al.}(2025)Huang, Liu, Bambi, Garc{\'i}a, \& Zhang}]{huangImpactReturningRadiation2025}
Huang, K., Liu, H., Bambi, C., Garc{\'i}a, J.~A., \& Zhang, Z. 2025, Physical Review D, 111, 063025, \dodoi{10.1103/PhysRevD.111.063025}

\bibitem[{Huang {et~al.}(2018)Huang, Qu, Zhang, Bu, Chen, Tao, Zhang, Lu, Li, Song, Xu, Cao, Chen, Liu, Chang, Yu, Weng, Hou, Kong, Xie, Zhang, ZHOU, Chang, Chen, Chen, Chen, Chen, Cui, Cui, Deng, Dong, Du, Fu, Gao, Gao, Gao, Ge, Gu, Guan, Gungor, Guo, Han, Hu, Huo, Ji, Jia, Jiang, Jiang, Jin, Jin, Li, Li, Li, Li, Li, Li, Li, Li, Li, Li, Li, Liang, Liao, Liu, Liu, Liu, Liu, Liu, Liu, Lu, Lu, Luo, Ma, Meng, Nang, Nie, Ou, Sai, Shang, Sun, Tan, Tao, Tuo, Wang, Wang, Wang, Wang, Wang, Wen, Wu, Wu, Xiao, Xiong, Xu, Yan, Yang, Yang, Yang, Zhang, Zhang, Zhang, Zhang, Zhang, Zhang, Zhang, Zhang, Zhang, Zhang, Zhang, Zhang, Zhang, Zhang, Zhang, Zhang, Zhang, Zhang, Zhao, Zhao, Zhao, Zheng, Zhu, Zhu, Zou, \& Collaboration}]{huangINSIGHTHXMTObservationsNew2018}
Huang, Y., Qu, J.~L., Zhang, S.~N., {et~al.} 2018, The Astrophysical Journal, 866, 122, \dodoi{10.3847/1538-4357/aade4c}

\bibitem[{Huppenkothen {et~al.}(2019)Huppenkothen, Bachetti, Stevens, Migliari, Balm, Hammad, Khan, Mishra, Rashid, Sharma, Ribeiro, \& Blanco}]{huppenkothenStingrayModernPython2019}
Huppenkothen, D., Bachetti, M., Stevens, A.~L., {et~al.} 2019, The Astrophysical Journal, 881, 39, \dodoi{10.3847/1538-4357/ab258d}

\bibitem[{Ingram \& Done(2011)}]{ingramPhysicalModelContinuum2011}
Ingram, A., \& Done, C. 2011, Monthly Notices of the Royal Astronomical Society, 415, 2323, \dodoi{10.1111/j.1365-2966.2011.18860.x}

\bibitem[{Ingram \& Done(2012)}]{ingramModellingVariabilityBlack2012}
---. 2012, Monthly Notices of the Royal Astronomical Society, 419, 2369, \dodoi{10.1111/j.1365-2966.2011.19885.x}

\bibitem[{Ingram {et~al.}(2009)Ingram, Done, \& Fragile}]{ingramLowfrequencyQuasiperiodicOscillations2009}
Ingram, A., Done, C., \& Fragile, P.~C. 2009, Monthly Notices of the Royal Astronomical Society: Letters, 397, L101, \dodoi{10.1111/j.1745-3933.2009.00693.x}

\bibitem[{Ingram {et~al.}(2019)Ingram, Mastroserio, Dauser, Hovenkamp, {van~der~Klis}, \& Garc{\'i}a}]{ingramPublicRelativisticTransfer2019}
Ingram, A., Mastroserio, G., Dauser, T., {et~al.} 2019, Monthly Notices of the Royal Astronomical Society, 488, 324, \dodoi{10.1093/mnras/stz1720}

\bibitem[{Ingram \& van~der Klis(2013)}]{ingramExactAnalyticTreatment2013}
Ingram, A., \& van~der Klis, M. 2013, Monthly Notices of the Royal Astronomical Society, 434, 1476, \dodoi{10.1093/mnras/stt1107}

\bibitem[{Ingram {et~al.}(2024)Ingram, Bollemeijer, Veledina, Dov{\v c}iak, Poutanen, Egron, Russell, Trushkin, Negro, Ratheesh, Capitanio, Connors, Neilsen, Kraus, Iacolina, Pellizzoni, Pilia, Carotenuto, Matt, Mastroserio, Kaaret, Bianchi, Garc{\'i}a, Bachetti, Wu, Costa, Ewing, Kravtsov, Krawczynski, Loktev, Marinucci, Marra, Miku{\v s}incov{\'a}, Nathan, Parra, Petrucci, Righini, Soffitta, Steiner, Svoboda, Tombesi, Tugliani, Ursini, Yang, Zane, Zhang, Agudo, Antonelli, Baldini, Baumgartner, Bellazzini, Bongiorno, Bonino, Brez, Bucciantini, Castellano, Cavazzuti, Chen, Ciprini, De~Rosa, Del~Monte, Di~Gesu, Di~Lalla, Di~Marco, Donnarumma, Doroshenko, Ehlert, Enoto, Evangelista, Fabiani, Ferrazzoli, Gunji, Hayashida, Heyl, Iwakiri, Jorstad, Karas, Kislat, Kitaguchi, Kolodziejczak, La~Monaca, Latronico, Liodakis, Maldera, Manfreda, Marin, Marscher, Marshall, Massaro, Mitsuishi, Mizuno, Muleri, Ng, O'Dell, Omodei, Oppedisano, Papitto, Pavlov, Peirson, Perri, {Pesce-Rollins}, Possenti, Puccetti, Ramsey,
  Rankin, Roberts, Romani, Sgr{\`o}, Slane, Spandre, Swartz, Tamagawa, Tavecchio, Taverna, Tawara, Tennant, Thomas, Trois, Tsygankov, Turolla, Vink, Weisskopf, Xie, \& Collaboration)}]{ingramTrackingXRayPolarization2024}
Ingram, A., Bollemeijer, N., Veledina, A., {et~al.} 2024, The Astrophysical Journal, 968, 76, \dodoi{10.3847/1538-4357/ad3faf}

\bibitem[{Ingram \& Motta(2019)}]{ingramReviewQuasiperiodicOscillations2019}
Ingram, A.~R., \& Motta, S.~E. 2019, New Astronomy Reviews, 85, 101524, \dodoi{10.1016/j.newar.2020.101524}

\bibitem[{Kalemci {et~al.}(2022)Kalemci, Kara, \& Tomsick}]{kalemciBlackHolesTiming2022}
Kalemci, E., Kara, E., \& Tomsick, J.~A. 2022, in Handbook of {{X-ray}} and {{Gamma-ray Astrophysics}}, ed. C.~Bambi \& A.~Santangelo (Singapore: Springer Nature), 1--43, \dodoi{10.1007/978-981-16-4544-0_100-1}

\bibitem[{Kara {et~al.}(2019)Kara, Steiner, Fabian, Cackett, Uttley, Remillard, Gendreau, Arzoumanian, Altamirano, Eikenberry, Enoto, Homan, Neilsen, \& Stevens}]{karaCoronaContractsBlackhole2019}
Kara, E., Steiner, J.~F., Fabian, A.~C., {et~al.} 2019, Nature, 565, 198, \dodoi{10.1038/s41586-018-0803-x}

\bibitem[{Kawamura {et~al.}(2023)Kawamura, Done, Axelsson, \& Takahashi}]{kawamuraMAXIJ1820+070Xray2023}
Kawamura, T., Done, C., Axelsson, M., \& Takahashi, T. 2023, Monthly Notices of the Royal Astronomical Society, 519, 4434, \dodoi{10.1093/mnras/stad014}

\bibitem[{Kotov {et~al.}(2001)Kotov, Churazov, \& Gilfanov}]{kotovXrayTimelagsBlack2001}
Kotov, O., Churazov, E., \& Gilfanov, M. 2001, Monthly Notices of the Royal Astronomical Society, 327, 799, \dodoi{10.1046/j.1365-8711.2001.04769.x}

\bibitem[{Kravtsov {et~al.}(2022)Kravtsov, Berdyugin, Kosenkov, Veledina, Piirola, Qadir, Berdyugina, Sakanoi, Kagitani, \& Poutanen}]{kravtsovOpticalPolarizationSignatures2022}
Kravtsov, V., Berdyugin, A.~V., Kosenkov, I.~A., {et~al.} 2022, Monthly Notices of the Royal Astronomical Society, 514, 2479, \dodoi{10.1093/mnras/stac1470}

\bibitem[{Krawczynski {et~al.}(2022)Krawczynski, Muleri, Dov{\v c}iak, Veledina, Rodriguez~Cavero, Svoboda, Ingram, Matt, Garcia, Loktev, Negro, Poutanen, Kitaguchi, Podgorn{\'y}, Rankin, Zhang, Berdyugin, Berdyugina, Bianchi, Blinov, Capitanio, Di~Lalla, Draghis, Fabiani, Kagitani, Kravtsov, Kiehlmann, Latronico, Lutovinov, Mandarakas, Marin, Marinucci, Miller, Mizuno, Molkov, Omodei, Petrucci, Ratheesh, Sakanoi, Semena, Skalidis, Soffitta, Tennant, Thalhammer, Tombesi, Weisskopf, Wilms, Zhang, Agudo, Antonelli, Bachetti, Baldini, Baumgartner, Bellazzini, Bongiorno, Bonino, Brez, Bucciantini, Castellano, Cavazzuti, Ciprini, Costa, De~Rosa, Del~Monte, Di~Gesu, Di~Marco, Donnarumma, Doroshenko, Ehlert, Enoto, Evangelista, Ferrazzoli, Gunji, Hayashida, Heyl, Iwakiri, Jorstad, Karas, Kolodziejczak, La~Monaca, Liodakis, Maldera, Manfreda, Marscher, Marshall, Mitsuishi, Ng, O'Dell, Oppedisano, Papitto, Pavlov, Peirson, Perri, {Pesce-Rollins}, Pilia, Possenti, Puccetti, Ramsey, Romani, Sgr{\`o}, Slane, Spandre,
  Tamagawa, Tavecchio, Taverna, Tawara, Thomas, Trois, Tsygankov, Turolla, Vink, Wu, Xie, \& Zane}]{krawczynskiPolarizedXraysConstrain2022}
Krawczynski, H., Muleri, F., Dov{\v c}iak, M., {et~al.} 2022, Science, 378, 650, \dodoi{10.1126/science.add5399}

\bibitem[{Kylafis {et~al.}(2008)Kylafis, Papadakis, Reig, Giannios, \& Pooley}]{kylafisJetModelGalactic2008}
Kylafis, N.~D., Papadakis, I.~E., Reig, P., Giannios, D., \& Pooley, G.~G. 2008, Astronomy \& Astrophysics, 489, 481, \dodoi{10.1051/0004-6361:20079159}

\bibitem[{Liu \& Qiao(2022)}]{liuAccretionBlackHoles2022}
Liu, B.~F., \& Qiao, E. 2022, iScience, 25, 103544, \dodoi{10.1016/j.isci.2021.103544}

\bibitem[{Lucchini {et~al.}(2023)Lucchini, Have, Wang, Homan, Kara, Adegoke, Connors, Dauser, Garcia, Mastroserio, Ingram, van~der Klis, K{\"o}nig, Lewin, Mallick, Nathan, O'Neill, Panagiotou, Piotrowska, \& Uttley}]{lucchiniVariabilityPredictorHardtosoft2023}
Lucchini, M., Have, M.~T., Wang, J., {et~al.} 2023, The Astrophysical Journal, 958, 153, \dodoi{10.3847/1538-4357/ad0294}

\bibitem[{Lyubarskii(1997)}]{lyubarskiiFlickerNoiseAccretion1997}
Lyubarskii, {\relax Yu}.~E. 1997, Monthly Notices of the Royal Astronomical Society, 292, 679, \dodoi{10.1093/mnras/292.3.679}

\bibitem[{Ma {et~al.}(2021)Ma, Tao, Zhang, Zhang, Bu, Ge, Chen, Qu, Zhang, Lu, Song, Yang, Yuan, Cai, Cao, Chang, Chen, Chen, Chen, Chen, Chen, Cui, Cui, Deng, Dong, Du, Fu, Gao, Gao, Gao, Gu, Guan, Guo, Han, Huang, Huo, Ji, Jia, Jiang, Jiang, Jin, Jin, Kong, Li, Li, Li, Li, Li, Li, Li, Li, Li, Li, Li, Liang, Liao, Liu, Liu, Liu, Liu, Liu, Liu, Lu, Lu, Luo, Luo, Meng, Nang, Nie, Ou, Sai, Shang, Song, Sun, Tan, Tuo, Wang, Wang, Wang, Wang, Wang, Wang, Wen, Wu, Wu, Wu, Xiao, Xiao, Xie, Xiong, Xu, Xu, Yang, Yang, Yang, Yi, Yin, You, Zhang, Zhang, Zhang, Zhang, Zhang, Zhang, Zhang, Zhang, Zhang, Zhang, Zhang, Zhang, Zhang, Zhang, Zhang, Zhang, Zhao, Zhao, Zheng, Zhou, Zhou, Zhu, Zhu, \& Zhuang}]{maDiscoveryOscillations2002021}
Ma, X., Tao, L., Zhang, S.-N., {et~al.} 2021, Nature Astronomy, 5, 94, \dodoi{10.1038/s41550-020-1192-2}

\bibitem[{Malzac {et~al.}(2001)Malzac, Beloborodov, \& Poutanen}]{malzacXraySpectraAccretion2001}
Malzac, J., Beloborodov, A.~M., \& Poutanen, J. 2001, Monthly Notices of the Royal Astronomical Society, 326, 417, \dodoi{10.1046/j.1365-8711.2001.04450.x}

\bibitem[{Mandal {et~al.}(2024)Mandal, Saha, Pal, \& Manna}]{mandalMultiwavelengthObservationMAXI2024}
Mandal, M., Saha, D., Pal, S., \& Manna, A. 2024, Astrophysics and Space Science, 369, 18, \dodoi{10.1007/s10509-024-04280-z}

\bibitem[{Mastroserio {et~al.}(2019)Mastroserio, Ingram, \& {van~der~Klis}}]{mastroserioXrayReverberationMass2019}
Mastroserio, G., Ingram, A., \& {van~der~Klis}, M. 2019, Monthly Notices of the Royal Astronomical Society, 488, 348, \dodoi{10.1093/mnras/stz1727}

\bibitem[{M{\'e}ndez {et~al.}(2022)M{\'e}ndez, Karpouzas, Garc{\'i}a, Zhang, Zhang, Belloni, \& Altamirano}]{mendezCouplingAccretingCorona2022}
M{\'e}ndez, M., Karpouzas, K., Garc{\'i}a, F., {et~al.} 2022, Nature Astronomy, 6, 577, \dodoi{10.1038/s41550-022-01617-y}

\bibitem[{Mizumoto {et~al.}(2018)Mizumoto, Done, Hagino, Ebisawa, Tsujimoto, \& Odaka}]{mizumotoXrayShorttimeLags2018}
Mizumoto, M., Done, C., Hagino, K., {et~al.} 2018, Monthly Notices of the Royal Astronomical Society, 478, 971, \dodoi{10.1093/mnras/sty1114}

\bibitem[{Novikov \& Thorne(1973)}]{novikovAstrophysicsBlackHoles1973}
Novikov, I.~D., \& Thorne, K.~S. 1973, Astrophysics of Black Holes., 343--450

\bibitem[{Peng {et~al.}(2024)Peng, Zhang, Shui, Zhang, Kong, Chen, Wang, Ji, Qu, Tao, Ge, Chang, Li, Li, Yu, \& Yan}]{pengNICERNuSTARInsightHXMT2024}
Peng, J.-Q., Zhang, S., Shui, Q.-C., {et~al.} 2024, The Astrophysical Journal Letters, 960, L17, \dodoi{10.3847/2041-8213/ad17ca}

\bibitem[{Podgorn{\'y} {et~al.}(2022)Podgorn{\'y}, Dov{\v c}iak, Marin, Goosmann, \& R{\'o}{\.z}a{\'n}ska}]{podgornySpectralPolarizationProperties2022}
Podgorn{\'y}, J., Dov{\v c}iak, M., Marin, F., Goosmann, R., \& R{\'o}{\.z}a{\'n}ska, A. 2022, Monthly Notices of the Royal Astronomical Society, 510, 4723, \dodoi{10.1093/mnras/stab3714}

\bibitem[{Poutanen {et~al.}(1997)Poutanen, Krolik, \& Ryde}]{poutanenNatureSpectralTransitions1997}
Poutanen, J., Krolik, J.~H., \& Ryde, F. 1997, Monthly Notices of the Royal Astronomical Society, 292, L21, \dodoi{10.48550/arXiv.astro-ph/9709007}

\bibitem[{Pozdnyakov {et~al.}(1983)Pozdnyakov, Sobol, \& Syunyaev}]{pozdnyakovComptonizationShapingXray1983}
Pozdnyakov, {\relax LA}., Sobol, {\relax IM}., \& Syunyaev, R.~A. 1983, Astrophysics and Space Physics Reviews, 2, 189

\bibitem[{Remillard \& McClintock(2006)}]{remillardXRayPropertiesBlackHole2006}
Remillard, R.~A., \& McClintock, J.~E. 2006, Annual Review of Astronomy and Astrophysics, 44, 49, \dodoi{10.1146/annurev.astro.44.051905.092532}

\bibitem[{Rybicki \& Lightman(1991)}]{rybickiRadiativeProcessesAstrophysics1991}
Rybicki, G.~B., \& Lightman, A.~P. 1991, Radiative Processes in Astrophysics (John Wiley \& Sons)

\bibitem[{Samimi {et~al.}(1979)Samimi, Share, Wood, Yentis, Meekins, Evans, Shulman, Byram, Chubb, \& Friedman}]{samimiGX3394NewBlack1979}
Samimi, J., Share, G.~H., Wood, K., {et~al.} 1979, Nature, 278, 434, \dodoi{10.1038/278434a0}

\bibitem[{Shakura \& Sunyaev(1973)}]{shakuraBlackHolesBinary1973}
Shakura, N.~I., \& Sunyaev, R.~A. 1973, Astronomy and Astrophysics, 24, 337

\bibitem[{Shui {et~al.}(2024)Shui, Zhang, Peng, Zhang, Chen, Ji, Kong, Feng, Yu, Wang, Chang, Yin, Qu, Tao, Ge, Zhang, \& Li}]{shuiPhaseresolvedSpectroscopyLowfrequency2024}
Shui, Q.-C., Zhang, S., Peng, J.-Q., {et~al.} 2024, The Astrophysical Journal, 973, 59, \dodoi{10.3847/1538-4357/ad676a}

\bibitem[{Sridhar {et~al.}(2025)Sridhar, Ripperda, Sironi, Davelaar, \& Beloborodov}]{sridharBulkMotionsBlack2025}
Sridhar, N., Ripperda, B., Sironi, L., Davelaar, J., \& Beloborodov, A.~M. 2025, The Astrophysical Journal, 979, 199, \dodoi{10.3847/1538-4357/ada385}

\bibitem[{Strohmayer {et~al.}(1996)Strohmayer, Zhang, Swank, Smale, Titarchuk, Day, \& Lee}]{strohmayerMillisecondXRayVariability1996}
Strohmayer, T.~E., Zhang, W., Swank, J.~H., {et~al.} 1996, The Astrophysical Journal, 469, L9, \dodoi{10.1086/310261}

\bibitem[{Timmer \& Koenig(1995)}]{timmerGeneratingPowerLaw1995}
Timmer, J., \& Koenig, M. 1995, Astronomy and Astrophysics, 300, 707

\bibitem[{Uttley {et~al.}(2014)Uttley, Cackett, Fabian, Kara, \& Wilkins}]{uttleyXrayReverberationAccreting2014}
Uttley, P., Cackett, E.~M., Fabian, A.~C., Kara, E., \& Wilkins, D.~R. 2014, The Astronomy and Astrophysics Review, 22, 72, \dodoi{10.1007/s00159-014-0072-0}

\bibitem[{Uttley \& Casella(2014)}]{uttleyMultiWavelengthVariability2014}
Uttley, P., \& Casella, P. 2014, Space Science Reviews, 183, 453, \dodoi{10.1007/s11214-014-0072-4}

\bibitem[{Uttley \& Malzac(2024)}]{uttleyLargeComplexXray2024}
Uttley, P., \& Malzac, J. 2024, Monthly Notices of the Royal Astronomical Society, stae2514, \dodoi{10.1093/mnras/stae2514}

\bibitem[{Uttley \& McHardy(2001)}]{uttleyFluxdependentAmplitudeBroadband2001}
Uttley, P., \& McHardy, I.~M. 2001, Monthly Notices of the Royal Astronomical Society, 323, L26, \dodoi{10.1046/j.1365-8711.2001.04496.x}

\bibitem[{{van der Klis}(2000)}]{vanderklisMillisecondOscillationsXray2000}
{van der Klis}, M. 2000, Annual Review of Astronomy and Astrophysics, 38, 717, \dodoi{10.1146/annurev.astro.38.1.717}

\bibitem[{Wang {et~al.}(2021)Wang, Mastroserio, Kara, Garc{\'i}a, Ingram, Connors, van~der Klis, Dauser, Steiner, Buisson, Homan, Lucchini, Fabian, Bright, Fender, Cackett, \& Remillard}]{wangDiskCoronaJet2021}
Wang, J., Mastroserio, G., Kara, E., {et~al.} 2021, The Astrophysical Journal Letters, 910, L3, \dodoi{10.3847/2041-8213/abec79}

\bibitem[{Wang {et~al.}(2022{\natexlab{a}})Wang, Kara, Lucchini, Ingram, van~der Klis, Mastroserio, Garc{\'i}a, Dauser, Connors, Fabian, Steiner, Remillard, Cackett, Uttley, \& Altamirano}]{wangNICERReverberationMachine2022}
Wang, J., Kara, E., Lucchini, M., {et~al.} 2022{\natexlab{a}}, The Astrophysical Journal, 930, 18, \dodoi{10.3847/1538-4357/ac6262}

\bibitem[{Wang {et~al.}(2022{\natexlab{b}})Wang, Kawai, Shidatsu, Murata, Hosokawa, Hanayama, Horiuchi, \& Morihana}]{wangMultiwavelengthStudiesXray2022}
Wang, S., Kawai, N., Shidatsu, M., {et~al.} 2022{\natexlab{b}}, Monthly Notices of the Royal Astronomical Society, 514, 5320, \dodoi{10.1093/mnras/stac1503}

\bibitem[{Wang {et~al.}(2024)Wang, Yan, Xie, Wang, \& Ma}]{wangAtypicalLowfrequencyQuasiperiodic2024}
Wang, X.-L., Yan, Z., Xie, F.-G., Wang, J.-F., \& Ma, R.-Y. 2024, The Astrophysical Journal, 969, 152, \dodoi{10.3847/1538-4357/ad58d1}

\bibitem[{Wang \& Zhang(2024)}]{wangRmsFluxSlope2024}
Wang, Y., \& Zhang, S.-N. 2024, The Astrophysical Journal, 962, 53, \dodoi{10.3847/1538-4357/ad1ab1}

\bibitem[{Wilkins {et~al.}(2016)Wilkins, Cackett, Fabian, \& Reynolds}]{wilkinsModellingXrayReverberation2016}
Wilkins, D.~R., Cackett, E.~M., Fabian, A.~C., \& Reynolds, C.~S. 2016, Monthly Notices of the Royal Astronomical Society, 458, 200, \dodoi{10.1093/mnras/stw276}

\bibitem[{Wilkins \& Fabian(2013)}]{wilkinsOriginLagSpectra2013}
Wilkins, D.~R., \& Fabian, A.~C. 2013, Monthly Notices of the Royal Astronomical Society, 430, 247, \dodoi{10.1093/mnras/sts591}

\bibitem[{Xiao {et~al.}(2019)Xiao, Lu, Ma, Ge, Yan, Li, Tuo, Zhang, Zhang, Liu, Zhou, Zhang, Bu, Cao, Jiang, Chen, Zhang, Lu, Chen, Qu, Song, Zhang, Zhuang, Shang, \& Jin}]{xiaoTimingAnalysisSwift2019}
Xiao, G.~C., Lu, Y., Ma, X., {et~al.} 2019, Journal of High Energy Astrophysics, 24, 30, \dodoi{10.1016/j.jheap.2019.09.005}

\bibitem[{You {et~al.}(2018)You, Bursa, \& {\.Z}ycki}]{youXRayQuasiperiodicOscillations2018}
You, B., Bursa, M., \& {\.Z}ycki, P.~T. 2018, The Astrophysical Journal, 858, 82, \dodoi{10.3847/1538-4357/aabd33}

\bibitem[{You {et~al.}(2016)You, Straub, Czerny, Sobolewska, R{\'o}{\.z}a{\'n}ska, Bursa, \& Dov{\v c}iak}]{youTESTINGWINDEXPLANATION2016}
You, B., Straub, O., Czerny, B., {et~al.} 2016, The Astrophysical Journal, 821, 104, \dodoi{10.3847/0004-637X/821/2/104}

\bibitem[{You {et~al.}(2020)You, T.~{\textbackslash}.Zycki, Ingram, Bursa, \& Wang}]{youXRayQuasiperiodicOscillations2020}
You, B., T.~{\textbackslash}.Zycki, P., Ingram, A., Bursa, M., \& Wang, W. 2020, The Astrophysical Journal, 897, 27, \dodoi{10.3847/1538-4357/ab9838}

\bibitem[{You {et~al.}(2021)You, Tuo, Li, Wang, Zhang, Zhang, Ge, Luo, Liu, Yuan, Dai, Liu, Qiao, Jin, Liu, Czerny, Wu, Bu, Cai, Cao, Chang, Chen, Chen, Chen, Chen, Chen, Chen, Cui, Cui, Deng, Dong, Du, Fu, Gao, Gao, Gao, Gu, Guan, Guo, Han, Huang, Huo, Jia, Jiang, Jiang, Jin, Jin, Kong, Li, Li, Li, Li, Li, Li, Li, Li, Li, Li, Li, Liang, Liao, Liu, Liu, Liu, Liu, Liu, Lu, Lu, Lu, Luo, Luo, Ma, Meng, Nang, Nie, Ou, Qu, Sai, Shang, Song, Song, Sun, Tan, Tao, Wang, Wang, Wang, Wang, Wang, Wang, Wen, Wu, Wu, Wu, Xiao, Xiao, Xiong, Xu, Yang, Yang, Yang, Yi, Yin, You, Zhang, Zhang, Zhang, Zhang, Zhang, Zhang, Zhang, Zhang, Zhang, Zhang, Zhang, Zhang, Zhang, Zhang, Zhang, Zhao, Zhao, Zheng, Zhou, Zhou, Zhu, \& Zhu}]{youInsightHXMTObservationsJetlike2021}
You, B., Tuo, Y., Li, C., {et~al.} 2021, Nature Communications, 12, 1025, \dodoi{10.1038/s41467-021-21169-5}

\bibitem[{You {et~al.}(2023)You, Cao, Yan, Hameury, Czerny, Wu, Xia, Sikora, Zhang, Du, \& Zycki}]{youObservationsBlackHole2023}
You, B., Cao, X., Yan, Z., {et~al.} 2023, Science, 381, 961, \dodoi{10.1126/science.abo4504}

\bibitem[{Yu {et~al.}(2023)Yu, Bu, Liu, Huang, Zhang, Yang, Qu, Zhang, Song, Zhang, Jia, Ma, Tao, Ge, Liu, Yan, Cao, Chang, Chen, Chen, Chen, Ding, Guan, Jin, Kong, Li, Li, Li, Li, Liao, Liu, Liu, Lu, Ma, Nie, Ren, Sai, Tan, Tuo, Wang, Wang, Wu, Xiao, Yin, You, Zhang, Zhang, Zhang, Zhao, Zheng, \& Zhou}]{yuSpectraltimingStudyInner2023}
Yu, W., Bu, Q.-C., Liu, H.-X., {et~al.} 2023, The Astrophysical Journal, 953, 191, \dodoi{10.3847/1538-4357/acd9a2}

\bibitem[{Zdziarski \& Gierli{\'n}ski(2004)}]{zdziarskiRadiativeProcessesSpectral2004}
Zdziarski, A.~A., \& Gierli{\'n}ski, M. 2004, Progress of Theoretical Physics Supplement, 155, 99, \dodoi{10.1143/PTPS.155.99}

\bibitem[{Zhang {et~al.}(2017)Zhang, Wang, M{\'e}ndez, Chen, Qu, Altamirano, \& Belloni}]{zhangEvolutionPhaseLags2017}
Zhang, L., Wang, Y., M{\'e}ndez, M., {et~al.} 2017, The Astrophysical Journal, 845, 143, \dodoi{10.3847/1538-4357/aa8138}

\bibitem[{Zhang {et~al.}(2019)Zhang, Dov{\v c}iak, \& Bursa}]{zhangConstrainingSizeCorona2019}
Zhang, W., Dov{\v c}iak, M., \& Bursa, M. 2019, The Astrophysical Journal, 875, 148, \dodoi{10.3847/1538-4357/ab1261}

\bibitem[{Zhang {et~al.}(2022)Zhang, M{\'e}ndez, Garc{\'i}a, Zhang, Karpouzas, Altamirano, Belloni, Qu, Zhang, Tao, Zhang, Huang, Kong, Ma, Yu, Rawat, \& Bellavita}]{zhangEvolutionCoronaMAXI2022}
Zhang, Y., M{\'e}ndez, M., Garc{\'i}a, F., {et~al.} 2022, Monthly Notices of the Royal Astronomical Society, 512, 2686, \dodoi{10.1093/mnras/stac690}

\end{thebibliography}
\end{document}